\documentstyle[12pt]{article}

\def\buildurel#1\under#2{\mathrel{\mathop{\kern0pt #2}\limits_{#1}}} 
\def\be{\begin{equation}}
\def\ee{\end{equation}}
\def\IP{\hbox{\rm I\kern -1.6pt{\rm P}}}
\def\IC{{\hbox{\rm C\kern-.58em{\raise.53ex\hbox{$\scriptscriptstyle|$}} 
    \kern-.55em{\raise.53ex\hbox{$\scriptscriptstyle|$}} }}}
\def\IN{\hbox{I\kern-.2em\hbox{N}}}
\def\IR{\hbox{\rm I\kern-.2em\hbox{\rm R}}}
\def\ZZ{\hbox{{\rm Z}\kern-.3em{\rm Z}}}
\def\IT{\hbox{\rm T\kern-.38em{\raise.415ex\hbox{$\scriptstyle|$}} }}
 
\begin{document}
 
\title{Stationary Nonequilibrium States in Boundary \\ 
Driven Hamiltonian Systems:  Shear Flow}
\author{N.I. Chernov
\\ Department of Mathematics\\
University of Alabama in Birmingham\\
Birmingham, AL 35294\\
E-mail: chernov@math.uab.edu
\and J.L. Lebowitz
\\Department of Mathematics\\
Rutgers University\\
New Brunswick, NJ 08903\\
E-mail:  lebowitz@math.rutgers.edu
}
\date{  }
\maketitle
 
\begin{abstract}
We investigate stationary nonequilibrium states of systems of particles
moving according to Hamiltonian dynamics with specified potentials.  The
systems are driven away from equilibrium by Maxwell demon ``reflection
rules'' at the walls.  These deterministic rules conserve energy but not
phase space volume, and the resulting global dynamics may or may not be
time reversible (or even invertible).  Using rules designed to simulate
moving walls we can obtain a stationary shear flow.  Assuming that for
macroscopic systems this flow satisfies the Navier-Stokes equations, we
compare the hydrodynamic entropy production with the average rate of phase
space volume compression.  We find that they are equal {\it when} the
velocity distribution of particles incident on the walls is a local
Maxwellian.  An argument for a general equality of this kind, based on the
assumption of local thermodynamic equilibrium, is given.  Molecular dynamic
simulations of hard disks in a channel produce a steady shear flow with the
predicted behavior.
\end{abstract}

\renewcommand{\theequation}{\arabic{section}.\arabic{equation}}
 
\section{Introduction}
 
Stationary nonequilibrium states (SNS) of macroscopic systems must be
maintained by external inputs at their boundaries.  Since a complete
microscopic description of such inputs is generally not feasible it is
necessary to represent them by some type of modeling.  However, unlike
systems in equilibrium, which maintain themselves without external inputs
and for which one can prove (when not inside a coexistence region of the
phase diagram) that bulk behavior is independent of the nature of the
boundary interactions, we do not know how different microscopic modeling of
boundary inputs, representing fluxes of matter, momentum or energy, affects
the resulting SNS.
 
What is observed experimentally is that in regimes close to equilibrium,
when the fluxes are small, the bulk macroscopic behavior is determined by
the unique solution of the hydrodynamic equations, with specified boundary
conditions on the hydrodynamical variables such as density, temperature and
fluid velocity [1].  The situation may change dramatically however as soon
as the driving forces become sufficiently large for this solution to lose
stability.  We can then have the formation of coherent structures, such as
rolls or hexagons, whose pattern is influenced by details of the boundary
conditions [2].
 
Even in the absence of hydrodynamic instabilities, e.g.\ in passive heat
conducting systems or fluids in regimes of laminar flow, the SNS generally
have {\it very} long range microscopic correlations, with slow power law
decay, which can be measured experimentally [3].  This raises the
possibility that even in regimes of hydrodynamic stability the modeling of
the boundary inputs may have global, albeit subtle, effects on the nature
of the SNS.  In fact it is known that, even in the near equilibrium regime,
different modelings of the external drives produce very different types of
microscopic measures of (what appears to be) the same macroscopic SNS.
Thus, stochastic drives, such as thermal boundaries in which the particles
acquire, following a collision with the walls, a specified Maxwellian
velocity distribution generally lead to stationary measures absolutely
continuous with respect to Liouville measure [4].  The same is true for
systems driven by collisions with some simple kinds of Hamiltonian infinite
particle reservoirs specified by a given distribution prior to the
collisions [5].  Deterministic thermostatting schemes, on the other hand,
yield measures singular with respect to Liouville measure [6--9].
 
It is quite possible, even likely, that these great differences in the
structure of the microscopic SNS (mSNS) do not have any significant effect
on the bulk properties of stable macroscopic SNS (MSNS).  This is what
happens for macroscopic equilibrium systems which can be described by a
variety of microscopic ensembles, e.g.\ the canonical or micro-canonical
[10].  Unlike equilibrium, however, it is far from clear at present how to
characterize essential global features of mSNS.  We do expect however that
{\it locally} such system will be close to an equilibrium state, at least
when the inputs are confined to the boundaries [11].  Indeed, as we shall
discuss further later, this property of local thermodynamic equilibrium
(LTE) holds the key to understanding SNS of macroscopic systems in the
hydrodynamic regime.
 
Questions regarding the nature of mSNS have recently come to the fore due
to the combination of computer and analytical investigations of SNS with
various deterministic thermostatting devices [6--10, 12-18].  The
simulations have shown that the stationary states produced by these drives
behave, at least as far as linear transport coefficients and other gross
properties of MSNS go, in reasonable accord with known experimental and
theoretical results.  In addition, the simulations have found unexpected
interesting microscopic structures in these singular measures, e.g.
pairing rules for the Lyapunov exponents, a formula for large fluctuations
in the phase space volume contraction rate, etc.
 
The rigorous mathematical analysis of such systems has confirmed some
aspects of the simulation results [9, 14].  This has led Gallavotti and
Cohen [15], see also [10, 16], to postulate what they call the ``chaotic
hypothesis'', based on Ruelle's principle for turbulence: ``{\it A many
particle system in a stationary state can be regarded, for the purpose of
computing macroscopic properties, as a smooth dynamical system with a
transitive Axiom A global attractor.  In the reversible case it can be
regarded, for the same purposes, as a smooth transitive Anosov system}.''
 
This hypothesis was shown by Gallavotti and Cohen to imply, for SNS
produced by reversible thermostatted dynamics, a formula for the
fluctuation in the phase space volume contraction rate in agreement with
the computer simulations [13] and to be generally consistent with known
results, at least when the driving is not too large [16], see also [17,
18].  It also implies, or even presupposes, a strong form of ``equivalence
of ensembles'', for SNS with specified macroscopic flows.  This is, as
already noted, certainly in accord with experience on SNS close to
equilibrium where it can be understood as an expression of the existence of
LTE.  It may however also be true more generally, at least in some form.
 
In the  present  paper we investigate  a new  class of models  in which the
microscopic  dynamics  in the  bulk  of   the  system are   Hamiltonian and
reflection at the boundaries are deterministic and energy conserving.  This
permits us  to define a  phase space flow  $\dot{X} = {\cal  F} (X)$, $X$ a
point in (a  fixed energy surface of)  the  system's phase space.   In this
respect our model  is similar to  the bulk thermostatting schemes mentioned
earlier [6--8].  Unlike those  schemes, however, which modify the equations
of   motion in the  bulk of  the fluid, something  which is computationally
useful but has no counterpart in real  physical systems, our model is fully
realistic away  from the  boundaries.  In  this respect  it is  similar  to
models in which  the  driving force  is a  ``boundary layer''  of reservoir
particles or  is given by stochastic  thermal  boundaries [4, 5,  19].  Our
combination of  realistic bulk dynamics   and deterministic boundary drives
offers  a simple model  of an  mSNS which can  be  investigated by means of
classical dynamical systems theory.   It will hopefully   lead to a  better
understanding of SNS representing real systems.
 
	Our main conclusion can be summarized as follows: Deterministic
boundary driven models, reversible or not, accurately represent the bulk
behavior of MSNS.  There is an equality in these models between phase-space
volume contraction, and hydrodynamic bulk entropy production for SNS of
macroscopic systems in LTE.  Plausible arguments of why this should be so
and of how to connect entropy production in nonequilibrium states generated
by different modelings of the inputs are given in section 8.  This is
preceded by a detailed description of analytic and computer simulation
results for specific models producing shear flow. A preliminary account of
this work is presented in [20].

\section{Description of Models}
 
To make the analysis as concrete as possible we shall consider here SNS
representing shear flow in a two dimensional system; we imagine this to be
the surface of a cylinder of height and perimeter $L$, or a square box with
periodic boundary conditions along the $x$ axis, i.e. we identify the left
and right sides at $x=\pm L/2$. On the top and bottom sides of the box,
$y=\pm L/2$, are rigid walls at which stand watchful Maxwell demons who
make the particles reflect according to rules satisfying the following two
conditions:

\noindent (i) the reflected velocity is determined  by the incoming one
and is energy conserving;

\noindent (ii) the particles at the top wall are driven 
to the right, and those at the bottom wall are symmetrically driven to the
left.  The purpose of rule (ii) is to imitate moving walls and thus produce
a shear flow in the bulk of the system.  The use of a two dimensional
system and of symmetric rules for top and bottom is for simplicity only and
the reader is free to imagine instead a three dimensional channel with
different reflection rules at top and bottom.
 
Since reflections at the walls preserve the particle speed, the reflection
rule can be defined in terms of angles. For a particle colliding with the
top wall, let $\varphi$ and $-\psi$ be the angles which the incoming and
outgoing velocities make with the positive $x$-axis, the direction of the
``wall velocity'', so that $0\leq\varphi,\psi\leq\pi$. At the bottom wall,
the angles are measured between the velocity vectors and the negative
$x$-axis. Then, any reflection rule is completely specified by a function
$\psi = f(\varphi,v)$, where $v$ stands for the speed of the particle.
 
For simplicity, we only study here functions independent 
of $v$, which are the same for both walls  so that 
$$
        \psi=f(\varphi),\ \ \ \ \ \ 
          0\leq\varphi,\psi\leq\pi
  \eqno(2.1)
$$
In particular, the identity function $f(\varphi) =\varphi$ corresponds to
elastic reflections, and $\psi = \pi - \varphi$ gives complete velocity
reversal reflections.
 
One particularly simple reflection rule is given by 
$$
\psi = \cases{\varphi, & $\varphi \leq \varphi_0$\cr
\pi - \varphi, & $\varphi > \varphi_0$\cr}\eqno(2.2)
$$
which, for $\pi/2 \leq \varphi_0 < \pi$ clearly satisfies (ii).  Under this
rule, particles moving `in the direction of the wall velocity', with
$\varphi \leq \varphi_0$, reflect elastically at the wall, while those
moving `opposite the wall velocity', $\varphi > \varphi_0$ will reflect
straight back, with the velocity vector reversed.  This rule is non
invertible, and will be discussed further in [18], see also [23]. Here we
shall focus our attention on two invertible rules that we found
particularly interesting and which we used in our molecular dynamics
simulations.  They are
$$
   \psi = (\pi+b)-[(\pi+b)^2 -\varphi(\varphi+2b)]^{1/2} 
       \label{b}\eqno(2.3)
$$
with $b \geq 0$ and 
$$
     \psi =  c \varphi 
       \label{c}\eqno(2.4)
$$
with $0<c\leq 1$. We call these the b-model and the c-model, 
respectively. 
 
The graph of the function (2.3) is a circular arc terminating at the points
$(0,0)$ and $(\pi,\pi)$ lying below the diagonal $\psi=\varphi$. This
function has the symmetry, $\varphi = f^{-1}(\psi) =
\pi -f(\pi - \psi)$, which makes the dynamics time-reversible. 
Time reversal symmetry which is present in Hamiltonian dynamics as well as
in the usual Gaussian thermostatted models [6--9], means that the system
retraces its trajectory backwards in time following a velocity reversal of
all the particles.  This symmetry plays an essential role in some of the
analysis in [15]. The c-model is not time-reversible.  Similarities and
differences between the b and c models are therefore of particular interest
in determining the range of universality present in SNS.  We shall in fact
see that the b and c models behave in a similar way.
 
\noindent
\section{The Stationary State}
 
Let us consider now the time evolution and the nature of the stationary
states we might expect with our b or c dynamics for a system of $N$
particles with total energy $E$; average particle and energy densities
$\bar n = N/L^2$, $\bar e = E/L^2$.  For the sake of concreteness imagine
the particles to be hard disks with unit mass and unit diameter so $E$ is
just their kinetic energy.  When $b = \infty$ or $c=1$ the boundary
conditions correspond to elastic reflections and so, for $N$ greater than
one, but less than some jamming value, we expect that, starting with any
initial measure absolutely continuous with respect to the Liouville measure
on the $H=E$ surface, the ensemble density will approach (weakly), as $t
\to \infty$, the uniform density on this surface, i.e.\ we expect the
micro-canonical ensemble of a system of hard disks with mixed periodic and
reflecting boundaries to be ergodic and mixing.  (There is actually, in
addition to the energy, also a conserved total $x$-momentum which we fix to
be zero and ignore).
 
When $b \ne \infty$ or $c \ne 1$ the rules will clearly produce a drift to
the right near the top wall and to the left near the bottom wall.  This
drift should produce, for $L$ large compared to the mean free path
$(\pi\bar{n})^{-1}$, an mSNS representing a system with a shear flow [19,
20].  On the microscopic level we expect now that any initial ensemble
density absolutely continuous with respect to the microcanonical ensemble
will evolve, as $t \to \infty$, to a stationary measure $\hat \mu$ whose
Hausdorff dimension\footnote{Here the Hausdorff dimension of the measure
$\hat{\mu}$ is, in accord with Young [21] and Pesin [22], the minimum of
the Hausdorff dimension of subsets of full measure.  (Other definitions of
Hausdorff dimensions of measures are sometimes used, which makes this
notion confusing.)  The Hausdorff dimension in the Young-Pesin version
coincides with the information dimension for systems with nonvanishing
Lyapunov exponents [22].  Note that the support of the measure $\hat{\mu}$
may or may not be the entire phase space.} is less than the dimension of
the energy surface [6--10].  Such behavior is proven for a single particle,
subjected to an external force, moving among a fixed periodic array of
scatterers [9].
 
We note that when both walls ``move'' in the {\it same} direction, the
$x$-component of the total momentum of the system is a monotone
non-decreasing function of the number of collisions with the walls.  Since
this is bounded above, the initial ensemble density must converge (at least
in some weak sense) to a measure whose support is on configurations in
which the particles all move parallel to the $x$-axis.  This situation is
rather pathological.  We expect our system with top and bottom walls moving
in opposite directions, to reach and stay in an LTE state at least when the
shear is not too large.  This is consistent with $\hat \mu$ being singular:
even with its Hausdorff dimension of being only a fraction of the dimension
of the energy surface [10].
 
Assuming that our system will indeed go, for $N$ large, $\bar n$ and $e$
fixed, to an MSNS representing a fluid in shear flow, as is indeed seen in
our computer simulations to be described later, we consider now briefly the
purely hydrodynamical description of such an MSNS.  This is given by the
stationary solution of the compressible Navier-Stokes equations for the
fluid velocity in the $x$-direction $u(y)$, the temperature $T(y)$ and
density $n(y)$, in a uniform channel of width $L$ in which top and bottom
walls move with velocities $\pm u_b$ in the $x$-direction, have the same
temperature $T_b$, and we impose a no slip boundary condition [1]
 
These equations, which are derived on the assumption of LTE [11], have the form [1, 6]
$$
{d \over dy} p(n,T) = 0,
$$
$$
{d \over dy}\left (\eta {du \over dy}\right ) = 0,
\eqno (3.1)
$$
$$
{d \over dy}\left (\kappa {dT \over dy}\right ) + 
  \eta\left ({du \over dy}\right )^2 = 0.
$$
\addtocounter{equation}{1}
Here $p(n,T)$ is the (local) equilibrium pressure of the system at constant
density $n(y)$ and temperature $T(y)$, $\eta(n,T)$ is the viscosity and
$\kappa(n,T)$ is the heat conductivity.  Equations (3.1) are to be solved
subject to the boundary conditions $u(\pm L/2) = \pm u_b, {}~~~ T(\pm L/2) =
T_b$ and fixed average particle density~~~ $L^{-1} \int_{-L/2}^{L/2} n(y)\,
dy = \bar n$.  This gives
$$
p(n,T) = p(n_0, T_0), ~~~du/dy = \Pi/\eta, ~~~
-\kappa\, dT/dy =  J(y) = \Pi
u(y), \eqno(3.2)
$$
where $n_0 = n(0)$, $T_0 = T(0)$, $\Pi$ is the {\it constant} $x$-momentum
flux in the negative $y$-direction, $J(y)$ is the heat flux in the positive
$y$-direction, and we have used the symmetry of the flow about $y=0$.
These equations can be solved once $p$, $\kappa$ and $\eta$ are given as
functions of $n$ and $T$.  The solution will be unique when the average
shear, $\gamma = 2u_b /L$, is small, see [24].
 
Eq. (3.2) can be integrated further to give
$$
u(y) = \Pi y\,\left [ {1 \over y} \int^y_0 {1 \over \eta} dy\right ], 
   ~~~{dT \over dy} =
-{1 \over 2}\, \left ( {\eta \over \kappa}\right )\, {d \over dy} u^2\eqno(3.3)
$$
so that
$$
\Pi = \bar \eta \gamma, ~~~~ 1/\bar \eta = {1 \over L} \int^L_0 
   {dy \over\eta(n,T)}.\eqno(3.4)
$$
For dilute gases, $\eta/\kappa$ is a constant independent of $n$ and $T$ in
which case 
$$
T(y) = T_0 - {1 \over 2}\left ({\eta \over \kappa}\right )\, u^2(y).\eqno(3.5)
$$

 In some cases the $y$-variation in $\eta$ and $\kappa$ are so small across
the channel that we have an essentially linear flow regime with,
$$
u(y) = \gamma y, ~~~~~~~~~~~~
T(y) = T_0 - (\eta / 2\kappa) \gamma^2 y^2, ~~~~ {\rm and} ~~~~~  \Pi = \eta
\gamma.\eqno (3.6)
$$
The constancy of $p$ together with the specified average density then
determines $n(y)$.
 
\section{Entropy Production in SNS}
 
Entropy plays a central role in determining the time evolution and (final)
equilibrium states of isolated macroscopic systems.  Its microscopic
interpretation as the log of the phase space volume of all micro-states
consistent with a specified macroscopic description, was well understood by
Boltzmann and the other ``founding fathers'' of statistical mechanics,
although there is still much fuzziness and outright confusion surrounding
the subject.  We refer the reader to [25] and references there for a
discussion.
 
The role of entropy and/or entropy production in SNS is also very important
although much less clear.  By their nature truly SNS cannot occur in an
isolated finite system evolving under Hamiltonian (or quantum)
dynamics---the only truly stationary macroscopic state for such a system
being the equilibrium one.  The situation can be different for {\it ab
initio} infinite systems [26], but we shall not discuss that here.  We
shall instead describe now various aspects of entropy production in the
simple SNS corresponding to stable shear flow in finite systems considered
here.  We will then discuss in section 8 the connection between them and
what they teach us.\medskip
 
\noindent
{\bf a)  Hydrodynamics}
 
\setcounter{equation}{1}
 
The {\it hydrodynamic} entropy production $\sigma$ per unit volume in our
stationary system is given by the Onsager form [17]; see in particular
chapter 14 in Balian [1] and section 2.2 in ref. [7]
$$
\sigma (y) = \frac{\Pi}{T}\,
{du \over dy} + J(y)\, {d \over dy}\left (\frac 1T\right ) 
 = \Pi {d \over dy} \left (\frac uT\right )\eqno(4.1)
$$
where we used (3.2) in the second equality.  The total hydrodynamic entropy
production $\bar{R}$ due to the dissipative fluxes in the steady state is
then
\begin{eqnarray}
\bar R &=&   \int_{\rm Volume} \sigma\, d{\bf r} = 
  \int_{\rm Surface} [\Pi u/T]\, ds =
  \int_{\rm Surface}\bar j_b/T\, ds \nonumber\\ 
  &=& \bar J_b/T_b = 2L^2\Pi\left ({u_b
\over L}\right )/T_b= L^2 \Pi \gamma /T_b
\label{(4.2)}
\end{eqnarray}
where $\bar j_b$ is the heat flux per unit length and $\bar J_b$ the total
flux to the walls and we have taken the channel to be of length $L$ with
periodic boundary conditions in the $x$-direction.
 
Eq. (4.2) is interpreted in the macroscopic formulation of irreversible
thermodynamics [1] as an equality, in the stationary state, between the
hydrodynamic entropy produced in the interior and the entropy carried by
the entropy flux, equal to $J_b/T_b$, to the walls of the container.  To
maintain such a steady state in an experimental situation requires external
forces acting on the walls to make them move with velocities $\pm u_b$.
The work done by these forces, $|\Pi u_b|$ per unit wall area, is converted
to heat in the bulk of the fluid by the viscosity and then absorbed by the
walls acting as infinite thermal reservoirs.  The steady state hydrodynamic
entropy production in the system, $\bar R$, is also carried to the walls by
this heat flux.  If we imagine the walls as ``equilibrium'' thermal bath at
temperature $T_b$ then $\bar R$ is equal to the rate of their entropy
increase $(dS_{\rm eq} = dU/T)$: note that we are assuming here that there
is no slip between the temperature of the fluid at the walls and the
temperature of the walls.
 
\noindent
{\bf b) Microscopic:  Stochastic Reservoirs}
 
The existence of macroscopic steady states, satisfying the compressible
Navier-Stokes equations (3.1), can be proven in suitable scaling limits, by
starting from the Boltzmann equation [24].  In such analysis the walls are
typically modeled by stochastic thermal boundaries; following collisions
with the walls particles have a Maxwellian velocity distribution with mean
$\pm u_b$ and temperature $T_b$.  Going beyond the mesoscopic description
given by the Boltzmann equation it is expected that such thermal boundaries
will produce similar SNS for general fluid systems which will be described,
on the microscopic level, by a stationary measure on the phase space having
a density, $\bar \mu$, absolutely continuous with respect to Liouville
measure [5].  In fact it is possible to show, for systems in contact with a
thermal reservoir at temperatures $T$, that the total ``ensemble entropy
production''
$$
{\cal R}(t) = \dot S_G(t) +
\langle \tilde J \rangle/T\eqno(4.3)
$$
is non-negative [5].  
  Here,
$$
S_G(t) \equiv S_G(\mu(X,t)) \equiv  - \int
\mu(X,t) \log \mu(X,t)\, dX,\eqno(4.4)
$$
is the system's Gibbs entropy, $dX$ is the Liouville volume element in the
phase space, and $\langle \tilde J\rangle$ is the ensemble average of a
phase space function $\tilde J(X)$ representing energy flux to the
reservoir.  At the same time the rate of change of the mean energy in the
system is given by
$$
{d \over dt} \int \mu H\, dx = \langle {\dot H} \rangle = -\langle \tilde J
\rangle + \langle \tilde W \rangle.\eqno (4.5)
$$  
  where $\langle \tilde W \rangle $ is the average mechanical work done on
the system by some external force, e.g.\ one produced by moving {\it rough}
walls of the system in a channel, c.f.\ [19].
 
In the stationary state obtained in the limit $t \to \infty$, $\mu = \bar
\mu$, so $\langle H \rangle$ and $S_G$ are constant with $\langle W \rangle
= \langle \tilde J \rangle$ and ${\cal {R}} = \bar{\cal R} = \langle \tilde
J \rangle /T = \langle W \rangle /T$.  Hence if we identify
$\langle \tilde J \rangle$ with $J_b$ then $\bar{\cal R}$ is equal to $\bar
R$ given in (4.2).  For a system in contact with only a thermal reservoir,
the stationary state is the equilibrium one and ${\cal R}(t) = d/dt[S_G -
\langle H \rangle/T] \to 0$ as $t \to \infty$, see [27] and section
8. \medskip
 
\noindent
{\bf c)  Microscopic:  Deterministically Driven Systems}
 
Let us turn now to our models where the flow is deterministic and
collisions with the boundaries conserve energy.  It is not clear at all a
priori what should now correspond at the microscopic level to the
hydrodynamic entropy production in our system.  Following the work in
refs. [7--9], we note that for a deterministic flow in the phase space the
rate of change of the systems Gibbs entropy defined in (4.4), is given by
$$
\dot S_G(t) = \int \mu(X,t)({\rm div}\, \dot X)\, dX =
-M(t),\eqno(4.6)
$$
This vanishes for an isolated system evolving according to Hamiltonian
dynamics for which ${\rm div} \dot X = 0$, but not for dynamics which does
not conserve Liouville volume.  Furthermore, since we expect the stationary
measure for our dynamics, $\hat \mu$, to be singular with respect to
Liouville measure, we will have $S_G(t) \buildurel {t \to \infty}\under
\longrightarrow -\infty$.  At the same time since the convergence of $\mu$
to $\hat \mu$ is in the weak sense, we might have a non vanishing limit
$$
\dot S_G(t) \to \int \hat \mu(X)({\rm div}\, \dot X)\, d X = -{\bar M} \eqno(4.7)
$$
We can then interpret $\bar M$, the average compression rate of phase space
volume per unit time as the ``measure entropy production'' in the
stationary state.  The existence and negativity of the limit in (4.7) was
proven in [9] for a simple bulk thermostatted model.  The non-negativity of
$\bar M$ for a stationary $\hat \mu$ is proven in a suitable general
setting by Ruelle [23].
 
The behavior of ${\dot S}_G (t)$ in the driven deterministic case is to be
contrasted with the case of stochastic reservoirs considered earlier where
the stationary measure has a smooth density with $\dot S_G(t) \to 0$ as $t
\to \infty$, while ${\cal R} (t) \to \bar {\cal R} = \bar R$, the positive
hydrodynamic entropy production in the MSNS.  Now in the bulk thermostatted
models, [7--9], the equations of motions are such that $\bar{M}$ is
automatically equal to the ensemble average of microscopic quantities which
can be identified with thermodynamic forces and fluxes such as appear in
the macroscopic entropy production.  This is not the case for the models
considered here.  We have no a priori prescription of $u$ or $T$ anywhere
in the system and phase space volume gets compressed only at collisions of
a particle with the wall: the bulk dynamics being Hamiltonian.  We
therefore need to investigate here the relationship, if any, between $\bar
M$ and $\bar R$ for our system.  Unfortunately, a direct computation, using
only the given dynamics, is totally out of reach of our present
mathematical abilities.  What we shall do instead in the next section is to
make some reasonable assumptions on the nature of the microscopic SNS in
the limit when our system becomes of macroscopic size.  It will turn out
that these assumptions, which are satisfied for a system in LTE, lead to an
equality between $\bar M$ and $\bar R$.  This will be checked and confirmed
by computer simulations in section 7.  We will then argue, in section 8,
that such an equality holds in general when the macroscopic system is in a
state of LTE.
 
{\it Remark}.  It might be feasible to carry out such a rigorous analysis
of our model within the context of the Boltzmann equation, in analogy to
what is done for stochastic walls in [24].  Numerical simulations on models
$b$ and $c$ using the Direct Simulation Monte Carlo method for simulating
the Boltzmann collision term inside the channel are now being carried out
[26].  The results appear consistent with those in section 7.
 
\section{Calculation of $M$ in the Hydrodynamic Regime}
 
To obtain the rate of compression $\bar M$ for our system, we observe that
our dynamics preserves phase space volume except at collisions of a
particle with a wall.  Since these collisions take place
``instantaneously'' we can compute the compression occuring at a single
collision---ignoring the rest of the particles.  The compression is then
just equal to the ratio of the ``outgoing'' one particle phase space volume
($dx^\prime\, dy^\prime\, dv^\prime_x\, dv^\prime_y)$ to the incoming one
($dx\, dy\, dv_x\, dv_y$) in a time interval $dt$ containing the collision.
A little thought shows that $dx^\prime = dx$ and $|dv^\prime_x\,
dv^\prime_y/dv_x\, dv_y| = |v^\prime\, dv^\prime\, d\psi|/|v\, dv\, d\phi|
= |d\psi/d\phi|$ where $\phi$ and $\psi$ are the incoming and outgoing
angles, related by $\psi = f(\phi)$.  Similarly, $|dy^\prime/dy| =
|v^\prime_y\, dt/v_y\, dt| = |\sin \psi / \sin \phi|$.  Hence in every
collision between a particle and the wall the phase space volume is changed
by a factor
$$
      \left | \frac{\sin\psi\, d\psi}{\sin\varphi\, d\varphi} \right |
       =\frac{\sin f(\varphi)}{\sin\varphi}|f'(\varphi)| = 
          \left |{d \cos f(\varphi) \over d \cos \varphi} \right | \eqno(5.1)
$$  
where $f$ defines the reflection rule (2.1). The phase volume will be
reduced or increased depending on whether (5.1) is smaller or larger than
unity.
 
The mean exponential rate of compression of the phase volume per unit time
is then given by
$$
       \bar M = -2 \langle  N_c
       \log [f'(\varphi)\sin f(\varphi)/\sin\varphi ]\rangle _{\hat
{\mu}}\eqno(5.2) 
$$
where $N_c(\varphi)$ is the flux of particles entering a collision with the
top wall at angle $\varphi$.  The factor 2 comes from summing over top and
bottom walls and the average is taken with respect to the stationary
measure $\hat\mu$.
 
An exact evaluation of $\bar M$ given in (5.2) is currently far beyond our
abilities.  To proceed further we assume now that in the hydrodynamic
regime corresponding to $L \gg l$, and $\gamma l \ll 1$, where $l \sim
(\pi\bar{n})^{-1}$ is the mean free path between particle-particle
collisions, the density $\rho(v_1,v_2)$, of the velocity vectors, ${\bf v}
= (v_1,v_2)$, of particles entering a collision with the walls, is
Maxwellian with the (to be determined) mean value $(\bar{v},0)$ and
temperature $T_w$ for the top wall, ($(-\bar{v},0)$ and $T_w$ for the
bottom wall).  That is
$$    \rho(v_1,v_2)=(2\pi T_w^3)^{-1/2} v_2\exp\left (-\frac{(v_1-\bar{v})^2
       +v_2^2}{2T_w}\right ), \ \ \ \ \ v_2>0\eqno(5.3)
$$
 
It will be convenient to rewrite the density (5.3) in polar coordinates,
$v_1=r\cos\theta$ and $v_2=r\sin\theta$,
$$    \rho(r,\theta)=\frac{1}{\sqrt{2\pi}\, T_w^{3/2}}\,
       r^2\sin\theta\cdot\exp\left (-\frac{\bar{v}^2
         +r^2-2\bar{v}r\cos\theta}{2T_w}\right ), ~~~~ 0 \leq \theta
\leq \pi\eqno(5.4)
$$
and denote by $\langle F\rangle$ the average of any function
$F(r,\theta)$  with respect to (5.4).
 
\setcounter{equation}{5}
 
The average momentum transfer `from wall to wall', $\hat{\Pi}$, is now
given by
$$
    \hat {\Pi}=n_c\left\langle \Delta v_1\right\rangle
       =n_c\left\langle v_{1,{\rm out}}-v_{1,{\rm in}}\right\rangle\eqno(5.5)
$$  
where $n_c$ is the `collision rate', i.e. the average number of collisions
with the top wall per unit time per unit length, which will have to be
computed later.  In our polar coordinates $(r,\theta)$ this value is
\begin{eqnarray}
  \hat{\Pi}&=&n_c\left\langle r(\cos f(\theta)-\cos\theta)\right\rangle\nonumber\\
       &=&\frac{n_c}{\sqrt{2\pi}\, T_w^{3/2}}
           \int_0^\infty\int_0^\pi (\cos f(\theta)-\cos\theta )
       r^3\sin\theta\nonumber\\
             &\times& \exp\left (-\frac{\bar{v}^2
         +r^2-2\bar{v}r\cos\theta}{2T_w}\right )\, d\theta dr \label{(5.6)}
\end{eqnarray}
 
In order to keep the ``velocity'' of our walls from growing with $L$ when
$L$ becomes macroscopic, which would certainly take the system away from
local equilibrium, we need to consider situations in which $\hat{\Pi}$,
like the hydrodynamic $\Pi$ in sec. 3 is of order $({1 \over L})$.  This
requires that the reflection rules (2.1) be close to the identity, i.e.\ we
put
$$
      \psi=f(\varphi)=\varphi +\delta f_1(\varphi)+\delta^2
        f_2(\varphi) + o(\delta^2)\eqno(5.7)
$$
with $f_1,f_2$ some fixed functions on $[0,\pi]$. In fact, as seen from
(5.6), $\delta$ has to be of order $O(L^{-1})$.  For our b-model (2.3) we
can set $\delta=1/b$, and then (5.7) will have the form:
$$
      \psi=\varphi - \delta\varphi(\pi-\varphi) + 
         \delta^2\varphi(\pi-\varphi)^2 + o(\delta^2)\eqno(5.8)
$$
For our c-model (2.4) we can set $\delta =1-c$, and so
$$
       \psi = \varphi - \delta\varphi\eqno(5.9)
$$
(Our calculations however are not restricted to these two models. )
 
\setcounter{equation}{9}
	
Expanding $\hat{\Pi}$ in $\delta$ gives 
\begin{eqnarray}
   \hat{\Pi} &=&n_c\left\langle r(-\sin \theta\cdot\delta f_1(\theta)
          +O(\delta^2))\right\rangle\nonumber\\
       &=&-\frac{n_c\delta}{\sqrt{2\pi}\, T_w^{3/2}}
           \int_0^\infty\int_0^\pi f_1(\theta)
       r^3\sin^2\theta\exp\left (-\frac{\bar{v}^2
         +r^2-2\bar{v}r\cos\theta}{2T_w}\right )\, d\theta dr\nonumber\\
       &+& O(\delta^2)
\label{(5.10)}
\end{eqnarray}
The last double integral depends on the so far unknown parameters $\bar{v}$
and $T_w$, and we denote it by $I(\bar{v},T_w)$.
 
We next assume that the steady state will indeed correspond to a shear flow
described (on the average) by the hydrodynamic equations in sec. 3, and
identify the $\hat \Pi$ with $\Pi$ and $T_w$ with $T_b$ there.  Since we
are however not given $T_b$ and $u_b$ they have to be determined from the
given data, assuming the system to be in LTE.  Thus the determination of
$T_w$ is now done on the basis that the system has a fixed energy $E$ so
that
$$
{1 \over L} \int^L_0 n(y)[T(y) + {1 \over 2} u^2(y)]dy = E/L^2 = \bar e,\eqno(5.11a)
$$
 
while the mean horizontal velocity of the particles near the wall can be
 taken as the mean between ingoing and outgoing velocities,
$$
u_b = \left\langle v_{1,{\rm out}}+v_{1,{\rm in}}\right\rangle /2 
= \bar v + O(\delta)\eqno(5.11b)
$$
The solution of (3.1) which determines $u(y), T(y)$ and $n(y)$ in terms of
$u_b$, $T_b$ and $\bar n$ will now be determined entirely by the a priori
given $\bar e, \bar n$ and the rule $f(\varphi)$, via (5.10) and (5.11).
The computation becomes straightforward in the linear approximation (3.6),
see eqs. (6.6)--(6.9) in the next section.

We study now the case $\delta\to 0$ and $L=a/\delta$ with some fixed $a>0$.
Then, the wall velocity $u_b$ is proportional to $\Pi ~L$ and thus does not
vanish as $\delta\to 0$.  Using (4.2) and (5.10) the hydrodynamic entropy
production, $\bar R$, is then, in this limit, given by
$$
     \bar R = -\frac{2an_c\bar{v}}{\sqrt{2\pi}\, T_w^{5/2}}
         I(\bar{v},T_w) + o(1)\eqno(5.12)
$$
 
\setcounter{equation}{12}
 
To obtain the compression rate $\bar M$ in this limit we expand (5.2) in
$\delta$ using our ansatz (5.3).  This gives, upon replacing $N_c$ by $2n_c
L$,
\begin{eqnarray}
\bar M/(2n_c L) &=& -\left\langle \log [f'(\varphi)\sin f(\varphi)/\sin\varphi ]\right\rangle
     \nonumber\\
      &=& -\left\langle f_1(\varphi)\cos\varphi/\sin\varphi\right\rangle\delta 
       - \left\langle f_1'(\varphi)\right\rangle\delta + o(\delta)
\end{eqnarray}
Now, integration by parts yields
$$
       \left\langle f_1'(\varphi)\right\rangle = - 
       \int_0^\infty\int_0^\pi f_1(\theta)\rho_{\theta}(r,\theta)
         \, d\theta dr
$$
where $\rho_\theta$ stands for the partial derivative of the 
density (5.4) with respect to $\theta$:
$$
       \rho_\theta(r,\theta) = \frac{\cos\theta}{\sin\theta}
       \rho(r,\theta) - \frac{\bar{v}r\sin\theta}{T_w}\rho(r,\theta)\eqno(5.14)
$$
Combining (5.13) and (5.14) then gives 
$$
\bar M = -2\delta Ln_c \bar{v}T_w^{-1}\left\langle f_1(\varphi)r\sin\varphi\right\rangle 
        + o(L\delta) = -\frac{2an_c\bar{v}}{\sqrt{2\pi}\, T_w^{5/2}}
         I(\bar{v},T_w) + o(1)\eqno(5.15)
$$
 
A shorter way to get (5.15) is to rewrite the middle in (5.13) as
$$
\left\langle \log \left (1 + {d[\cos f(\varphi) - \cos \varphi] \over d \cos
\varphi} \right )\right\rangle 
= -\delta \left\langle  (\sin \varphi) {df_1 \over d \cos \varphi} \right\rangle +
o(\delta)\eqno(5.16)
$$
and then do an integration by parts using $\cos \theta$ as a variable.  The
leading term in (5.16) will be recognized as corresponding to (4.6) for the
continuous time action of the thermostats, see Appendix.
 
The leading term in the expansion of $\bar M$ is thus exactly the same as
in that of $\bar R$, hence $\bar M$ and $\bar R$ become equal in the
hydrodynamical limit, $L \to \infty$.  The essential requirement for the
equality is the validity of (5.14).  (This is weaker than (5.3), permitting
multiplication of the Maxwellian there by an arbitrary function of $r$, but
we do not know of any physically reasonable non Maxwellian $\rho$ which
would satisfy (5.14).) The equality between $\bar R$ and $\bar M$ in the
hydrodynamical regime is thus a nontrivial consequence of our (local
equilibrium) assumption (5.3).  As already noted this is an important
difference between our models and the bulk thermostatted models of [7, 8,
9].  The relation between phase-space volume compression and what {\it
looks like} entropy production is so built into the structure of the
dynamics of the latter that there equality holds, essentially by
definition, even for systems consisting of just of one or a few particles
which are certainly not in local equilibrium.  This is not the case here.
The equality fails for systems with too few particles to be well described
by hydrodynamics, as is seen in the next section.

\section{Small $\delta$ regime}
\label{secSLR}
\setcounter{equation}{0}
 
Before presenting our numerical simulations, which were obviously done at
finite (and not so large) $L$, we consider the consequences which can be
drawn from our assumption (5.3) when $\delta\to 0$ and $L$ is (relatively)
large but held fixed. Now the generated shear flow in our system will only
be approximately described by the hydrodynamics equation (3.1).
Furthermore $u_w$ as well as $\bar R$ and $\bar M$ vanish as $\delta\to 0$.
Interestingly enough there is now a difference between the b and c models
with ${\bar M}/\bar R$ remaining finite for the b-model as $\delta \to 0$
(close to unity for moderately large $L$), while becoming infinite for the
c-model.  As we shall see, however, this is connected with the details of
the c-model, rather than with its lack of time reversibility.
 
To compute $\bar R$ and $\bar M$ in this case it will be necessary to
expand various quantities up to second order in $\delta$.  Thus,
\begin{eqnarray}
{\bar M}/(2n_c L) &=& \left \langle \log [f'(\varphi)\sin f(\varphi)/\sin\varphi ]\right\rangle\nonumber\\
      &=& \left\langle f_1(\varphi)\cos\varphi/\sin\varphi\right\rangle\delta 
       + \left\langle f_1'(\varphi)\right\rangle\delta\nonumber\\
      &+&\left\langle f_2(\varphi)\cos\varphi/\sin\varphi\right\rangle\delta^2 
       + \left\langle f_2'(\varphi)\right\rangle\delta^2\nonumber\\
      &-&\frac 12 \left\langle (f_1'(\varphi))^2\right\rangle\delta^2 
       - \frac 12 \left\langle f_1^2(\varphi)/\sin^2\varphi\right\rangle\delta^2 
         + o(\delta^2)
\end{eqnarray}
This expansion is of course meaningful only if the coefficients of $\delta$
and $\delta^2$ are finite which requires that $f_1(0)=f_1(\pi)=0$ and that
$f_1$ has finite derivatives at both $0$ and $\pi$.  Our b-model satisfies
these assumption, because $f_1(\varphi)=-\varphi(\pi-\varphi)$, see (5.8).
The computation of $\bar M$ for the c-model which does not satisfy
$f_1(\pi) = 0$ will be done separately.
 
In order to compute the coefficients in the expansion (6.1) we need to
expand the density (5.4) which using (3.2) or (3.6) depends on $\delta$
through the quantities $\bar{v} = O(\delta)$, $T_w=1/2+O(\delta^2)$ to
first order in $\delta$.  This gives
$$
    \rho(r,\theta)=2\pi^{-1/2} r^2e^{-r^2}\sin\theta 
         + 4 \pi^{-1/2}\bar v r^3e^{-r^2}\sin\theta\cos\theta 
             +o(\delta),\eqno(6.2)
$$
$$
{\bar M} = (C_1 + C_2)\bar v^2 + o(\delta^2) 
\eqno(6.3)
$$
with 
$$
      C_1 = 8\eta\left (1-\frac{\eta}{n_c L}\right )^{-2}\eqno(6.4)
$$
$$
     C_2 = \frac{2\pi\eta^2}{n_cL}
      \left (1-\frac{\eta}{n_c L}\right )^{-1}
      \left (\int_0^\pi f_1(\theta)\sin^2\theta\, d\theta\right )^{-2}
        \cdot \int_0^\pi [f_1^2(\theta)\sin^{-1}\theta 
           + (f_1'(\theta))^2\sin\theta]\, d\theta \eqno(6.5)
$$
 
\setcounter{equation}{6}
 
We assume now that for small $\delta$ and moderate $L$, of the kind used in
our simulation, the system is reasonably well described by a linear shear
flow, Eq. (3.6).  We can then find the relation between $\bar v$ and
$\delta$,
$$
\bar v = -{Ln_c \delta \over 2\eta \sqrt \pi} 
      \left (1-\frac{\eta}{n_c L}\right )\cdot
      \int_0^\pi f_1(\theta) \sin^2
       \theta\, d\theta + O(\delta^2)\eqno(6.6)
$$
Similarly we get
\begin{eqnarray}
       \Pi &=& -\frac{2n_c\delta}{\sqrt{\pi}}
           \int_0^\infty\int_0^\pi f_1(\theta)
       r^3\sin^2\theta\cdot e^{-r^2}\, d\theta dr + O(\delta^2)\nonumber\\
           &=& -\frac{n_c\delta}{\sqrt{\pi}}
           \int_0^\pi f_1(\theta)\sin^2\theta\, d\theta 
           + O(\delta^2)
         \label{6.7}
\end{eqnarray}
and
$$
  \bar R = \frac{2L\Pi v_w}{T_w} = \frac{2L^2n_c^2\delta^2}{\pi\eta}
         \left (1-\frac{\eta}{n_cL}\right )^{-1}\cdot
         \left (\int_0^\pi f_1(\theta)\sin^2\theta\, d\theta \right)^2 
              + o(\delta^2)\eqno(6.8)
$$
\setcounter{equation}{8}
The expression for $\bar M$ in (6.3) in terms of $\delta$ and $L$ is
\begin{eqnarray}
{\bar M} &=& \frac{2L^2n_c^2\delta^2}{\pi\eta}
      \left (\int_0^\pi f_1(\theta)\sin^2\theta\, d\theta\right )^2\nonumber\\
      &+& \frac 12 Ln_c\delta^2
      \left (1-\frac{\eta}{n_c L}\right )\cdot
         \int_0^\pi [f_1^2(\theta)\sin^{-1}\theta 
           + (f_1'(\theta))^2\sin\theta]\, d\theta 
              + o(\delta^2)
          \label{6.9}
\end{eqnarray}
 
Comparing (6.8) to (6.9) we see that the term of $\bar M$ proportional to
$L^2$ coincides with the corresponding term of $\bar R$ while the terms of
order $L$ differ.  The difference between $\bar M$ and $\bar R$ thus gets
relatively small as $L$ increases with the value of $\bar M$ remaining
slightly larger than the value of $\bar R$ as $\delta \to 0$.  This is in
agreement with the computer simulations.
 
\setcounter{equation}{9}
 
We now calculate $\bar M$ for our c-model where, according to (5.9),
$f_1(\varphi)=-\varphi$ and $f_2(\varphi)\equiv 0$.  The expansion (6.1)
now has a singular term and so we need to use the original formula (5.2)
\begin{eqnarray}
  {\bar M} &=& -2Ln_c \Big (\ln(1-\delta) + \left\langle 
              \ln\sin ((1-\delta)\varphi) 
          \right\rangle - \left\langle \ln\sin\varphi 
              \right\rangle \Big )\nonumber\\
        &=& -2Ln_c(I_0 + I_1 - I_2)
         \label{6.10}
\end{eqnarray}
where $I_0=-\delta+O(\delta^2)$ and the two other terms involve integration
with respect to the density (6.2).  It will be sufficient to integrate only
with the zeroth order term there since corrections will, as before, produce
terms of order $\delta^2$, which we can now disregard compared to the
leading term. We find
$$
    I_2 = \left\langle\ln\sin\varphi\right\rangle 
        = \frac 12 \int_0^\pi\sin\theta\,\ln\sin\theta\, d\theta 
        = \ln 2-1\eqno(6.11)
$$
and 
$$
    I_1 = \left\langle\ln\sin((1-\delta)\varphi)\right\rangle 
        = \frac 12 \int_0^\pi\sin\theta\,\ln\sin((1-\delta)\theta)\, d\theta \eqno(6.12)
$$
This yields:
$$
    I_1 = \ln 2 - 1 + \delta +\frac{\pi^2}{4}\delta^2\ln\delta 
        + O(\delta^2)\eqno(6.13)
$$
Combining the expansions for $I_0,I_1$ and $I_2$ gives 
$$
   {\bar M} = -2^{-1}\pi^2Ln_c \delta^2\ln\delta + O(\delta^2)\eqno(6.14)
$$
The corresponding expression for $\bar R$ has no singular terms
$$
\bar R = \frac{\pi^3L^2n_c^2}{8\eta}
      \left (1-\frac{\eta}{n_c L}\right )
             \delta^2   + o(\delta^2)\eqno(6.15)
$$
This gives
$$
\bar M/ \bar R = {\rm const}\cdot\ln(1/\delta) + O(1) \to\infty\eqno(6.16)
$$
In other words, for our c-model, when $L$ is finite and $\delta\to 0$, 
the quantities $\bar M$ and $\bar R$ have different rates of convergence 
to zero!\medskip
 
{\it Remark}. We also calculated the second order term in the 
expansion of $\bar M$:
$$
    {\bar M} = -2^{-1}\pi^2Ln_c \delta^2\ln\delta + C_2\delta^2 +
O(\delta^3)\eqno(6.17) 
$$
where 
$$
   C_2=\frac{\pi^3n_c^2L^2}{8\eta}-11.9202\, n_cL\eqno(6.18)
$$
 
\section{Numerical results}
 
Computer simulations of our b and c models were performed with $N=100$ and
$N=200$ hard disks of unit mass and diameter $\sigma = 1$.  We kept the
volume fraction occupied by the disks, $\pi \bar n/4$, equal to 0.1.  Our
system was thus in the dilute gas phase.  For $N=100$ this corresponds to
$L=28.0$ and for $N=200$ to $L = 39.6$. The mean free path $l$ at this
density is about 2.3 [27] so $L/l \sim 17$ for the latter.  (Discrepancies
between results of the simulations, and of hydrodynamics, can therefore be
expected, a priori, to be of order $(2u_b/17)$.)
 
The positions of the particle centers were chosen randomly (without
overlap) in a domain $|x|\leq L/2$, $|y| \leq (L-1)/2$.  The initial
velocities were generated randomly with subsequent normalization of the
total kinetic energy so that $2E = \sum v_i^2 = N$.
 
There is an obvious instability in the dynamics, and the round-off errors
accumulate at an exponential rate. In fact, the system loses its memory
completely after every 100-200 collisions in the box.  The dynamical
meaning of computer simulations of unstable dynamical systems like ours is
a difficult question, which has been discussed recently in the literature
[28]. One way to think of round-off errors is as small perturbations of the
deterministic trajectory of the phase point made at every collision. So,
instead of a true trajectory we can only track a perturbed one, or a
pseudo-trajectory.  Then the question is -- what are the quantities
measured by averaging along such pseudo-trajectories?
 
One possible answer is given by the so-called shadowing lemma [29].  It
says that for smooth hyperbolic systems every pseudo-trajectory is shadowed
(approximated) by a real trajectory. The distance between the two
trajectories is of the order of the computer accuracy. This may justify
averaging over computer-generated pseudo-trajectories for smooth hyperbolic
systems although it certainly does not guarantee that these finite time
averages represent typical behavior [28].  Moreover, our dynamics are not
smooth, and hyperbolicity cannot be easily established.  Trust in the
results of our simulations, like all others done on such unstable dynamical
systems therefore relies mainly on faith in some kind of typicality
resulting from the (pseudo) random effects of the roundoffs [28].
 
In any case to prevent the system from leaving the energy surface due to
round-off errors, the total kinetic energy was renormalized after every $N$
collisions in the box.
 
With the above reservations we believe that the results to be described are
statistically reliable within one percent. This is based on various checks
comparing different runs and different levels of computer precision.  For
each value of $b$ and $c$ we averaged over about 25,000 collisions per
particle with other particles and about 1,200 with the walls. We also
changed the computer precision from single (7 accurate decimals) to double
(14 decimals) and the so called long double (19 decimals) to make sure the
results were stable.  Most of the programming was done in the C language on
an 80486/DX-66 PC and on a SUN SPARC station 1000 at the University of
Alabama at Birmingham.
 
In each run the vertical height, occupied by the centers, $L-1$, was
divided into twenty equally spaced horizontal layers and time averages of
the density $n(y)$, mean $x$-velocity $u(y)$ and variances
$\langle(v_x-u)^2\rangle, \langle v_y^2
\rangle$ were taken.  We also recorded time averages of $x$-momentum
transfer from the walls, $\Pi$, and the compression rate $M$.
 
Figures 1 and 2 show typical velocity and temperature profiles $u(y)$ for
the b- and c-models with $N=200$ particles.  The velocity profiles are
almost linear and the temperature profiles $T(y) = {1 \over 2} \langle (v_x
- u(y))^2 + v^2_y \rangle$ almost quadratic, away from the walls,
consistent with the approximation (3.6).  The deviations from linearity
near the walls is due both to the dependence of $\eta$ and $\kappa$ on $T$
and $n$ which means that we should use (3.3) rather use (3.6).  Since
$\eta(n,T) \sim \sqrt T$ this will indeed lead to increases in the slope
near the wall.  There are of course also effects due to the deviations from
the hydrodynamic limit of $O(2u_b l/L)$, but these are harder to compute.
\setlength{\unitlength}{0.02in}
 
\begin{figure}
\begin{picture}(300,180)(0,0)
\put(100,10){\line(0,1){160}}
\put(200,10){\line(0,1){160}}
\put(50,90){\line(1,0){200}}
\multiput(50,88)(10,0){21}{\line(0,1){5}}
\put(45,81){-0.5}
\put(147,81){0.5}
\put(145,94){-0.5}
\put(247,94){0.5}
\put(110,160){$b=10$}
\put(210,160){$c=0.9$}
\put(246,81){$u(y)$}
\put(94,160){$y$}
\put(194,160){$y$}
\put(41.5,14){\circle*{2}}
\put(49.7,22){\circle*{2}}
\put(56.6,30){\circle*{2}}
\put(62.7,38){\circle*{2}}
\put(68.4,46){\circle*{2}}
\put(74.0,54){\circle*{2}}
\put(80.0,62){\circle*{2}}
\put(86.0,70){\circle*{2}}
\put(91.5,78){\circle*{2}}
\put(97.2,86){\circle*{2}}
\put(102.8,94){\circle*{2}}
\put(108.3,102){\circle*{2}}
\put(114.1,110){\circle*{2}}
\put(119.6,118){\circle*{2}}
\put(125.4,126){\circle*{2}}
\put(131.4,134){\circle*{2}}
\put(137.4,142){\circle*{2}}
\put(143.8,150){\circle*{2}}
\put(150.1,158){\circle*{2}}
\put(158.3,166){\circle*{2}}
\put(155.3,14){\circle*{2}}
\put(161.9,22){\circle*{2}}
\put(166.8,30){\circle*{2}}
\put(171.2,38){\circle*{2}}
\put(175.7,46){\circle*{2}}
\put(180.2,54){\circle*{2}}
\put(184.7,62){\circle*{2}}
\put(188.9,70){\circle*{2}}
\put(193.2,78){\circle*{2}}
\put(197.9,86){\circle*{2}}
\put(202.3,94){\circle*{2}}
\put(206.5,102){\circle*{2}}
\put(211.0,110){\circle*{2}}
\put(215.2,118){\circle*{2}}
\put(219.6,126){\circle*{2}}
\put(224.4,134){\circle*{2}}
\put(229.1,142){\circle*{2}}
\put(233.5,150){\circle*{2}}
\put(238.5,158){\circle*{2}}
\put(245.0,166){\circle*{2}}
\end{picture}
\caption{Velocity profiles $u(y)$ for the models $b=10$ 
and $c=0.9$.}
\end{figure}
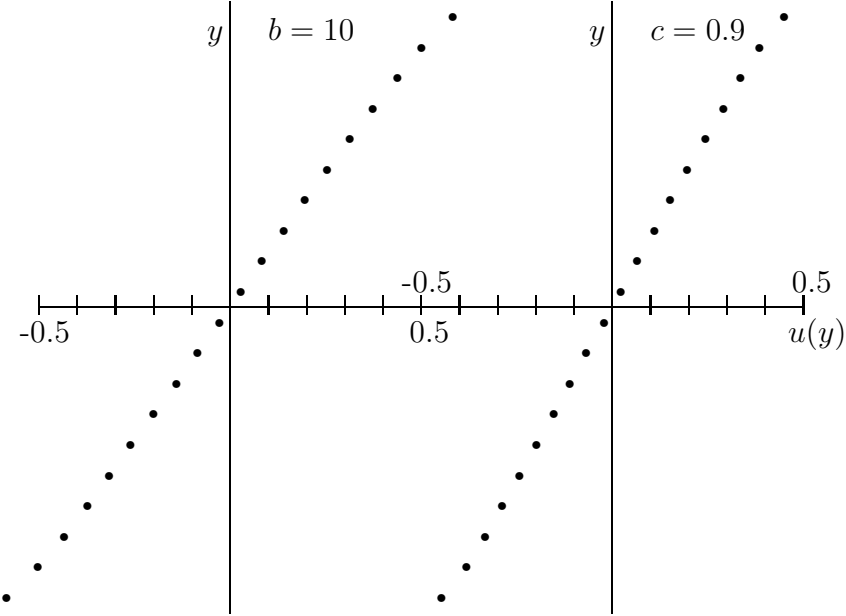
 
\begin{figure}
\begin{picture}(300,180)(0,0)
\put(100,10){\line(0,1){160}}
\put(200,10){\line(0,1){160}}
\put(50,90){\line(1,0){210}}
\multiput(50,88)(10,0){22}{\line(0,1){5}}
\put(45,81){-0.5}
\put(147,81){0.5}
\put(247,81){0.5}
\put(257,94){$T(y)$}
\put(110,130){$b=10$}
\put(210,130){$c=0.9$}
\put(94,160){$y$}
\put(194,160){$y$}
\put(140.0,14){\circle*{2}}
\put(142.3,22){\circle*{2}}
\put(143.2,30){\circle*{2}}
\put(144.0,38){\circle*{2}}
\put(144.5,46){\circle*{2}}
\put(145.0,54){\circle*{2}}
\put(145.5,62){\circle*{2}}
\put(145.6,70){\circle*{2}}
\put(145.7,78){\circle*{2}}
\put(145.8,86){\circle*{2}}
\put(145.9,94){\circle*{2}}
\put(145.7,102){\circle*{2}}
\put(145.5,110){\circle*{2}}
\put(145.3,118){\circle*{2}}
\put(144.9,126){\circle*{2}}
\put(144.2,134){\circle*{2}}
\put(143.7,142){\circle*{2}}
\put(142.9,150){\circle*{2}}
\put(141.8,158){\circle*{2}}
\put(139.8,166){\circle*{2}}
\put(243.8,14){\circle*{2}}
\put(245.1,22){\circle*{2}}
\put(245.8,30){\circle*{2}}
\put(246.3,38){\circle*{2}}
\put(246.8,46){\circle*{2}}
\put(247.0,54){\circle*{2}}
\put(247.3,62){\circle*{2}}
\put(247.5,70){\circle*{2}}
\put(247.5,78){\circle*{2}}
\put(247.6,86){\circle*{2}}
\put(247.6,94){\circle*{2}}
\put(247.4,102){\circle*{2}}
\put(247.4,110){\circle*{2}}
\put(247.4,118){\circle*{2}}
\put(247.3,126){\circle*{2}}
\put(247.0,134){\circle*{2}}
\put(246.6,142){\circle*{2}}
\put(246.1,150){\circle*{2}}
\put(245.5,158){\circle*{2}}
\put(244.1,166){\circle*{2}}
\end{picture}
\caption{Temperature profiles $T(y)$ for the models $b=10$ 
and $c=0.9$.}
\end{figure}
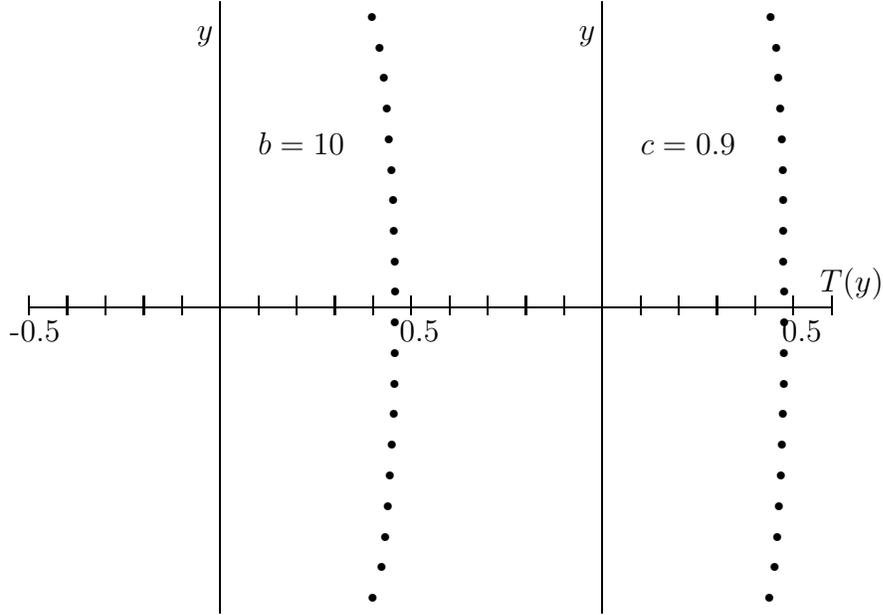
 
We estimated the shear rate $d u(y)/dy$, by the least square fit of a line
$u(y)=\gamma y$ to the experimental velocity profile and used the
experimental momentum transfer from wall to wall, $\Pi$, to find the
average shear viscosity as defined in (3.4), $\bar \eta =\Pi/\gamma$.  To
mitigate the problem arising from the nonlinearity of the profile near the
wall we used only the data in the bulk of the system, from level 4 to level
17, i.e. we excluded the top three and the bottom three levels. Table~1
presents the experimental values of the shear viscosity computed for the b-
and c-dynamics with $N=200$ particles.  The essentially linear dependence
of $\Pi$ on $\delta$ is certainly consistent with (6.6).
 
\begin{center}
\begin{tabular} {||c||c|c|c||}\hline\hline
model & $\Pi_{\rm exp}$ & $\eta_{\rm exp}$ & Enskog $\eta$ \\ \hline\hline
b=10 &  $6.03\times 10^{-3}$ & 0.204 & 0.209 \\
b=25 &  $3.12\times 10^{-3}$ & 0.213 & 0.219 \\ 
b=50 &  $1.66\times 10^{-3}$ & 0.221 & 0.222 \\ 
b=100 &  $0.85\times 10^{-3}$ & 0.218 & 0.222 \\ 
b=200 &  $0.43\times 10^{-3}$ & 0.216 & 0.222 \\ \hline
c=0.90 &  $4.87\times 10^{-3}$ & 0.213 & 0.214 \\ 
c=0.95 &  $2.82\times 10^{-3}$ & 0.221 & 0.220 \\ 
c=0.97 &  $1.77\times 10^{-3}$ & 0.206 & 0.221 \\ 
c=0.99 &  $0.62\times 10^{-3}$ & 0.229 & 0.222 \\ \hline\hline
\end{tabular}
\end{center}
 
\begin{center}
Table 1. Computed and theoretical values of $\eta$. 
\end{center} 
 
The shear viscosity of hard disk fluids can be estimated via Enskog's
modification of the Boltzmann equation [30, 31]. It gives the value
$$
    \eta_E=\eta_{\rm dilute}
    \left [\chi^{-1}+bn+0.8729\chi(bn)^2\right ]\eqno(7.1)
$$
where $\eta_{\rm dilute}$ is the value obtained from the Boltzmann equation
[30] 
$$
    \eta_{\rm dilute} = 1.022\cdot \frac{1}{2} 
     \sqrt{\frac{kT}{\pi}}\eqno(7.2)
$$
for disks of unit mass and diameter.  In (7.1) $b$ is the second virial
coefficient, $b=\pi/2$, and $\chi$ is the Enskog scaling factor which is
just the equilibrium pair-correlation function at contact. We have
estimated $\chi$ by using the pressure equation for hard disk fluids,
$$
    \frac{p}{nkT}=1+\frac{\pi}{2}n\chi\eqno(7.3)
$$
and its virial expansion in the number density $n$, or equivalently the
``scaled particle'' approximation see [33, 34] 
$$
    \frac{p}{nkT}=1+bn+b_3n^2+b_4n^3+\ldots
        \approx \left [ 1 - {\pi n \over 4} \right ]^{-2}\eqno(7.4)
$$

In our case $bn=0.2$, and we get
$$
    \eta_E \approx 1.081\, \eta_{\rm dilute}\eqno(7.5)
$$
 
The value of $\eta_E$ so computed using the measured mean temperature and
density in rows 4--17 is shown in the last column of Table~1. The agreement
is very good as would be expected for the low density system we are
considering.
 
We next checked the constancy of the pressure $p(n,T)$.  To do this it is
important to note that Eq. (7.3) is valid, even for uniform equilibrium
systems, {\it only} in the bulk, i.e.\ at distances from the wall large
compared to the equilibrium correlation length.  For our dilute gas system
this length is of order $\sigma = 1$.  Near the wall the density becomes
nonuniform with the density {\it at the wall}, $n_w$, equal to $p/kT$; $T$
the uniform equilibrium temperature [30].  Since the hydrodynamic
eqs. (3.1) are valid on a scale which is very large compared to any
microscopic scale this variation in density is not considered there.  The
situation is very different however for our computer simulations where we
can expect to see these density variations, c.f.\ [19].  Indeed, the
pressure defined as the average $y$-momentum transfer from the top wall in
the $y$-direction per unit length and unit time would be using (5.3), in
analogy to (5.6), equal to $n_c \langle r(\sin f(\theta) + \sin
\theta)\rangle$.  This would reduce to $n_w kT$ to zero order in $\delta$.
 
We present in Table 2 experimental values of $n$, $T$, as well as the bulk
pressure defined in (7.3) in the different layers taking the mean of the
values in the layers situated symmetrically about the middle. Here the
density is given by the average number of particles on each level.  Since
$n(y)$ increases rapidly as we approach the wall, we do not have a simple
formula for $p$ in the top column.  Using $n_w = p/kT$ leads to an
extrapolated density at the wall of about 13.4 and 13 for the $b=10$ and
$c=.9$ cases.
 
\begin{center}
\begin{tabular} {||r||c|c|c|c||c|c|c|c||}\hline\hline
      & \multicolumn{4}{c||}{$b=10$} & \multicolumn{4}{c||}{$c=0.9$} \\ \cline{2-9}
level & $v_x$ & $T$ & $n$ & $p$ & $v_x$ & $T$ & $n$ & $p$ \\ \hline\hline
1 &  0.584 & 0.399 & 11.35 & ~~~~~ & 0.449 & 0.439 & 10.70 & ~~~~~ \\
2 &  0.502 & 0.420 & 10.35 & 5.330 & 0.383 & 0.453 & 10.15 & 5.722 \\ 
3 &  0.437 & 0.430 & 10.13 & 5.338 & 0.333 & 0.459 & 10.06 & 5.724 \\ 
4 &  0.374 & 0.438 &  9.97 & 5.337 & 0.289 & 0.465 &  9.94 & 5.713 \\
5 &  0.315 & 0.443 &  9.86 & 5.331 & 0.244 & 0.469 &  9.89 & 5.712 \\ 
6 &  0.257 & 0.449 &  9.77 & 5.341 & 0.197 & 0.471 &  9.85 & 5.713 \\ 
7 &  0.198 & 0.454 &  9.67 & 5.341 & 0.153 & 0.473 &  9.84 & 5.715 \\
8 &  0.141 & 0.455 &  9.67 & 5.342 & 0.111 & 0.474 &  9.82 & 5.720 \\ 
9 &  0.084 & 0.457 &  9.63 & 5.344 & 0.067 & 0.475 &  9.82 & 5.718 \\ 
10 &  0.028 & 0.459 &  9.60 & 5.340 & 0.022 & 0.476 & 9.80 & 5.714 \\ \hline\hline
\end{tabular}
\end{center}
 
\begin{center}
Table 2. Experimental measurements of the $x$-velocity, temperature $T$,
density $n$ and pressure $p$ on all levels for the $b=10$ and $c=0.9$
models.
\end{center}

  The agreement between the prediction of the hydrodynamic equation (3.6)
and the simulation results shows that the Maxwell demon boundary drives
indeed set up, in the limit $L \to \infty$, an MSNS for shear flow.  We
therefore computed the values of $\bar R$ and $\bar M$ defined for our
shear flow by formulas (5.2) and (4.2) with $\Pi$ given by (5.5)
experimentally, by using time averages of the appropriate dynamical
functions in our simulations.  We also computed these quantities
numerically according to our integral formulas using the Maxwellian ansatz
(5.4). In both these computations we used the experimental values of $T_w$,
$v_w$, and $n_c$. The results are presented in Table 3.  The last column of
Table 3 gives the leading term in $\delta$, computed from (5.12), for which
$\bar M = \bar R$.

\begin{center}
\begin{tabular} {||c||c|c|c||}\hline\hline
model & $M_{\rm th}/M_{\rm exp}$ & $R_{\rm th}/R_{\rm exp}$  & 
$M_{\rm lead}$ \\ \hline\hline
b=10 &  0.761/0.740  & 0.695/0.767 & 0.867 \\
b=25 &  0.161/0.157  & 0.156/0.162 & 0.169 \\ 
b=50 &  0.0429/0.0417  & 0.0422/0.0428 & 0.0437 \\ 
b=100 & 0.0113/0.0110 & 0.0111/0.0112 & 0.0113 \\ 
b=200 & 0.00294/0.00285 & 0.00289/0.00289 & 0.00291 \\ \hline
c=0.9 & 0.448/0.444 & 0.405/0.432 & 0.402 \\ 
c=0.95 & 0.149/0.148 & 0.127/0.131 & 0.126 \\ 
c=0.97 & 0.0641/0.0632 & 0.0523/0.0531 & 0.0518 \\ 
c=0.99 & 0.00871/0.00868 & 0.00628/0.00632 & 0.00625 \\ \hline\hline
\end{tabular}
\end{center}
 
\begin{center}
Table 3. The experimental and theoretical values of $M$ and $R$. 
\end{center} 
 
The agreement between the so computed theoretical values of $\bar M$ and
$\bar R$ and their experimental ones is quite good. This suggests that the
integral formulas (5.6) and (5.2) with the Maxwellian density (5.4) are
quite accurate.

We also tested directly this hypothesis using a chi-square test and the
Kolmogorov-Smirnov test, see, e.g., [36]. Both tests accepted the
distribution (5.3) for the velocity vectors of incoming particle colliding
with the wall for various values of $b$ and $c$. Just to determine the
sensitivity of our procedure, we tested in the same way the distribution of
the {\em outgoing} velocity vectors (whose Maxwellianity would contradict
our assumption (5.3)).  assumed, and which would have contradicted our
assumption (5.3)).  As expected, both tests frequently rejected this last
hypothesis for several values of $b$ and $c$, indicating that the
reflection rule (2.1) distorts the velocity distribution in a considerable
way even for small $\delta$.
 
\noindent  {\it Remarks}
 
1) Our analysis in section 6 shows that the quantities $\bar M$ and $\bar
R$ are of order $v_w^2$ as $\delta\to 0$, with the exception of $\bar M$
for the c-model, which is of order $v_w^2\ln(1/v_w)$. Figure~3 shows the
ratios $M/v_w^2$ and $R/v_w^2$ as functions of $\delta$. For the b-model
they both `nicely' converge to the same positive constant ($\approx 1.6$)
as $\delta\to 0$. For the c-model the ratio $R/v_w^2$ also converges to a
number ($\approx 1.55$), while $M/v_w^2$ apparently grows to infinity, in
agreement with the prediction in sec. 6.

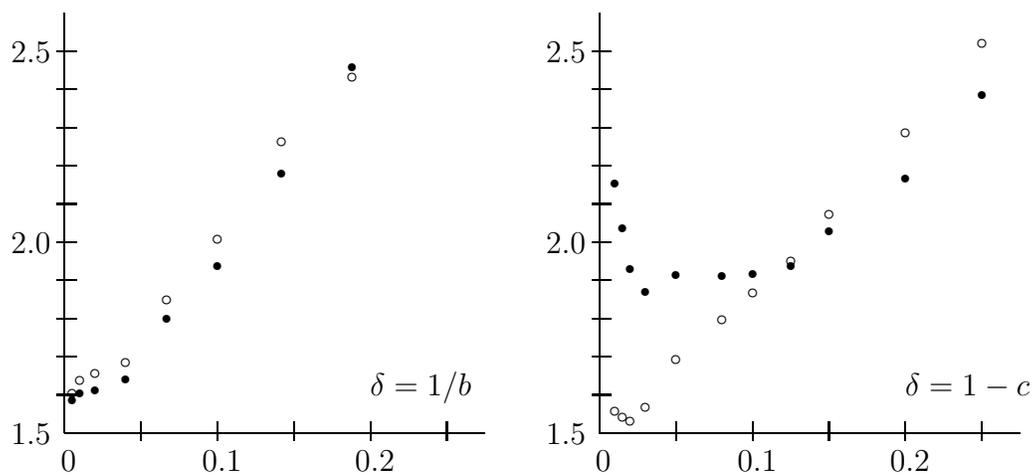
\begin{figure}
\begin{picture}(300,150)(0,0)
\put(30,20){\line(0,1){110}}
\put(30,20){\line(1,0){110}}
\multiput(50,18)(20,0){5}{\line(0,1){5}}
\multiput(28,30)(0,10){10}{\line(1,0){5}}
\put(170,20){\line(0,1){110}}
\put(170,20){\line(1,0){110}}
\multiput(190,18)(20,0){5}{\line(0,1){5}}
\multiput(168,30)(0,10){10}{\line(1,0){5}}
\put(29,10){0}
\put(66,10){0.1}
\put(106,10){0.2}
\put(169,10){0}
\put(206,10){0.1}
\put(246,10){0.2}
\put(110,30){$\delta=1/b$}
\put(250,30){$\delta=1-c$}
\put(16,17){1.5}
\put(16,67){2.0}
\put(16,117){2.5}
\put(156,17){1.5}
\put(156,67){2.0}
\put(156,117){2.5}
 
\put(105.2,115.8){\circle*{2}}  
\put(105.2,113.1){\circle{2}}
\put(86.7,87.9){\circle*{2}}   
\put(86.7,96.3){\circle{2}}
\put(70,63.7){\circle*{2}}  
\put(70,70.9){\circle{2}}
\put(56.7,50.0){\circle*{2}}
\put(56.7,55.0){\circle{2}}
\put(46,34.1){\circle*{2}}  
\put(46,38.6){\circle{2}}
\put(38,31.3){\circle*{2}}  
\put(38,35.7){\circle{2}}
\put(34,30.4){\circle*{2}}  
\put(34,33.7){\circle{2}}
\put(32,28.5){\circle*{2}}  
\put(32,30.5){\circle{2}}

\put(270,108.4){\circle*{2}}
\put(270,122.0){\circle{2}}
\put(250,86.8){\circle*{2}} 
\put(250,98.5){\circle{2}}
\put(230,72.8){\circle*{2}} 
\put(230,77.2){\circle{2}}
\put(220,63.9){\circle*{2}} 
\put(220,65.1){\circle{2}}
\put(210,61.7){\circle*{2}} 
\put(210,56.6){\circle{2}}
\put(202,61.0){\circle*{2}} 
\put(202,49.6){\circle{2}}
\put(190,61.4){\circle*{2}} 
\put(190,39.3){\circle{2}}
\put(182,57.0){\circle*{2}} 
\put(182,26.9){\circle{2}}
\put(178,63.0){\circle*{2}} 
\put(178,23.2){\circle{2}}
\put(176,73.7){\circle*{2}} 
\put(176,24.3){\circle{2}}
\put(174,85.3){\circle*{2}} 
\put(174,25.8){\circle{2}}

\end{picture}
\caption{The ratios $M/v_w^2$ (solid circles) and 
$R/v_w^2$ (hollow circles) versus $\delta=1/b$ and $\delta=1-c$.}
\end{figure}

2)  An interesting question is how the velocity near the 
wall $v_w$ depends on the model parameter $\delta$ 
as the size $L$ is hold fixed and $\delta$ is not very small.   
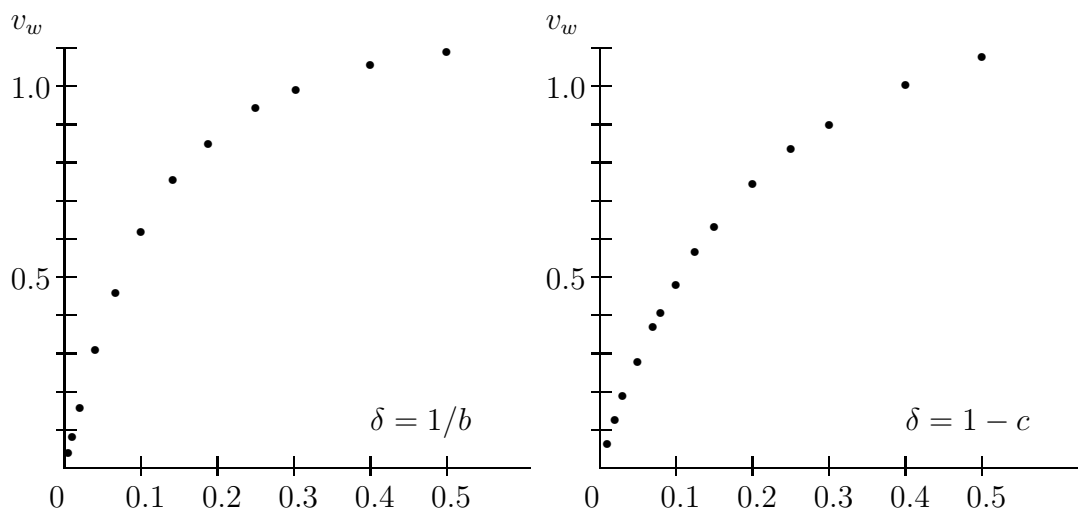
\begin{figure}
\begin{picture}(300,150)(0,0)
\put(30,20){\line(0,1){110}}
\put(30,20){\line(1,0){122}}
\multiput(50,18)(20,0){5}{\line(0,1){5}}
\multiput(28,30)(0,10){11}{\line(1,0){5}}
\put(170,20){\line(0,1){110}}
\put(170,20){\line(1,0){125}}
\multiput(190,18)(20,0){5}{\line(0,1){5}}
\multiput(168,30)(0,10){11}{\line(1,0){5}}
\put(26,10){0}
\put(46,10){0.1}
\put(66,10){0.2}
\put(86,10){0.3}
\put(106,10){0.4}
\put(126,10){0.5}
\put(166,10){0}
\put(186,10){0.1}
\put(206,10){0.2}
\put(226,10){0.3}
\put(246,10){0.4}
\put(266,10){0.5}
\put(110,30){$\delta=1/b$}
\put(250,30){$\delta=1-c$}
\put(16,67){0.5}
\put(16,117){1.0}
\put(156,67){0.5}
\put(156,117){1.0}
\put(16,135){$v_w$}
\put(156,135){$v_w$}
 
\put(130,129.1){\circle*{2}}  
\put(110,125.6){\circle*{2}}  
\put(90.6,119.1){\circle*{2}} 
\put(80,114.4){\circle*{2}}   
\put(67.6,105.0){\circle*{2}} 
\put(58.3,95.4){\circle*{2}}  
\put(50,82.0){\circle*{2}}    
\put(43.3,65.9){\circle*{2}}  
\put(38,51.0){\circle*{2}}    
\put(34,35.7){\circle*{2}}    
\put(32,28.3){\circle*{2}}    
\put(31,24.1){\circle*{2}}

\put(270,127.8){\circle*{2}}
\put(250,120.5){\circle*{2}}
\put(230,110.0){\circle*{2}}
\put(220,103.6){\circle*{2}}
\put(210,94.3){\circle*{2}} 
\put(200,83.1){\circle*{2}} 
\put(195,76.7){\circle*{2}} 
\put(190,68.1){\circle*{2}} 
\put(186,60.7){\circle*{2}} 
\put(184,57.0){\circle*{2}} 
\put(180,47.7){\circle*{2}} 
\put(176,38.8){\circle*{2}} 
\put(174,32.7){\circle*{2}} 
\put(172,26.4){\circle*{2}}

\end{picture}
\caption{The velocity $v_w$ near the wall versus 
$\delta=1/b$ and $\delta=1-c$.}
\end{figure}
Fig.~4 shows the experimental values of $v_w$ versus $\delta$ as $\delta$
varies between zero and .5.  A linear regression for small $\delta$ is very
clear for both models. For large $\delta$, the function $v_w(\delta)$
increases more and more slowly, and apparently has a finite asymptote
$v_w(+\infty)$. This happens for a simple reason -- the energy balance in
the system. Given a linear velocity profile $u(y)=\gamma y$ the energy
balance with the minimal temperature $T (y)\equiv 0$ gives $\gamma_{\max} =
\sqrt{6}/L$, so that
$$
         v_{w,\max} = \sqrt{6}/2 \approx 1.225\eqno(7.6)
$$
Indeed, the largest velocity near the wall we observed in our simulations
(say, with $b=0$ or $c=0.2$) was around 1.2.  Under these conditions,
however, the laminar velocity profile breaks down.

\vfill \eject 
\section{Discussion}
 
In Sec. 5 we obtained an equality between the phase volume contraction
$\bar M$ defined in (5.2) for the mSNS produced by our model and the
hydrodynamic entropy production $\bar R$ defined in (4.2) for the MSNS, in
the limit $L \to \infty$,~~ $\delta \to 0$, with $\delta L = a$~~ fixed.
To understand the origin of this equality, also found approximately in our
computer simulations, we will analyze here in more detail, the production
of entropy in nonequilibrium macroscopic systems, discussed in Sec. 4 for
SNS.  This will make use of formal manipulations of various expressions for
the entropy of such systems whose justification requires, at the minimum,
the validity of dissipative hydrodynamics, e.g.\ the Navier-Stokes eqs. of
section 3, obtained as a scaling limit in going from microscopic to
macroscopic descriptions of our system.  It will assume {\it ipso facto}
the existence of LTE in these systems since this is required for the
derivation of the hydrodynamic equations [11].  While these assumptions are
very reasonable for systems not too far from equilibrium, where the
interactions (collisions) between the particles, tending to bring the
system to equilibrium, dominate locally over the external forces and fluxes
pushing the system away from equilibrium, such results are {\it very far}
from being proven for systems with Hamiltonian dynamics.  Even their
derivation from the Boltzmann eq. is still incomplete at the present time
[24].  Given this situation there seems little point in giving any proofs
here---even for those parts where this may be possible.  Instead the
analysis should be thought of as heuristic and suggestive.
 
For a system in LTE with hydrodynamical variables $n({\bf r},t), {\bf
u}({\bf r},t), e({\bf r},t)$, corresponding to particle density, velocity,
and energy density, evolving according to hydrodynamical equations, the
hydrodynamic entropy, $S_h$, of the system at any time $t$ is the integral
of $s_{\rm eq}(n,e^\prime)$, the entropy density in a {\it uniform}
equilibrium system with densities $n$ and $e^\prime$, [1, 11, 25],
$$
S_h \equiv \int_{\rm Volume} s_{\rm eq} (n({\bf r}), e^\prime({\bf r}))\, d{\bf r}\eqno(8.1)
$$
where $e^\prime({\bf r}) \equiv [e({\bf r}) - {1 \over 2} n({\bf r}) {\bf
u}^2(r)]$ is the thermal energy density.  For an equilibrium system, ${\bf
u}$ has to be independent of ${\bf r}$ and can therefore be removed by a
Gallilean transformation.
 
We also have, essentially from the definition of LTE, that macroscopic
systems in LTE have a {\it local} microscopic description which is, to
leading order, the same as that for a {\it uniform} equilibrium system with
the same parameters; i.e.\ if we consider a ``small'' {\it macroscopic
volume element} around ${\bf r}$, its properties will be approximately
given by the grand-canonical ensemble, assumed to be equivalent to the
canonical or micro-canonical ensemble, specified by the local values of
$n$, $\bf u$, $e$, at time $t$ [11, 34].  The dissipative fluxes are then
related by the transport coefficients to the gradients of the
hydrodynamical variables, which are very small on the micro scale.
 
Let us call $\cal M$ the macroscopic state specified by the hydrodynamical
variables $\{n, {\bf u}, e\}$ and let $\nu(X; {\cal M})$ be the grand
canonical ensemble with {\it local} chemical potential and temperature
appropriate to $n({\bf r}), e^\prime({\bf r})$. Then we have that $S_h$ is
equal to the Gibbs entropy $S_G$ defined in (4.4), corresponding to $\mu =
\nu$,
$$
S_h \simeq S_G(\nu) \equiv -\int \nu(X; {\cal M}) \log \nu(X; {\cal M})\, dX,\eqno(8.2)
$$
see [11, 25, 34].  
 
It is furthermore true, as observed by Boltzmann, that the hydrodynamic
entropy $S_h$, of a system in the macrostate $\cal M$, is equal to the
logarithm of the phase space volume, $\Gamma(\cal M)$, associated with the
macroscopic state ${\cal M} = \{ n, {\bf u}, e\}$, see [25], i.e.
$$
S_h \simeq S_B({\cal M}) \equiv  \log~ \Gamma({\cal M})\eqno(8.3)
$$
(Note that $S_B({\cal M})$ may make sense even when $\cal M$ does not
correspond to an LTE state, when $S_h$ is not well defined.  We shall
however not consider such cases here, so (8.3) will always hold.)
 
The equality between the different expressions for the macroscopic entropy
of systems in LTE given in (8.1), (8.2) and (8.3), depends crucially on the
large separation of scales between the micro and macro descriptions. It
holds to leading order in the ratio of micro to macro scales, e.g.\ $l/L$
defined in sec. 7, and becomes exact, in the sense that it has the same
limiting value when divided by the number of particles in the system, only
when the ratio of micro to macro scales goes to zero, the hydrodynamical
scaling limit.  For real systems the equality is only approximate, as are
the hydrodynamic eqs., etc.  We shall however, following Newton in a
different but similar context [35], assume that ``the error will not be
sensible; and therefore this ... may be considered {\it as physically
exact}'' (italics added) and will therefore treat (8.2) and (8.3) as true
equalities.
 
Taking the time derivative of (8.1) using the standard equilibrium
relations for derivatives of $s_{eq}$, we obtain
$$
\dot S_h = \int_{\rm Volume} \left\{-{\lambda \over T} {\partial n \over
\partial t} + {1 \over T}\left [{\partial e \over \partial t} - {\partial \over
\partial t}\left ({1 \over 2} nu^2\right )\right ]\right \}d{\bf r}, \eqno(8.4)
$$
where $\lambda({\bf r},t)$ is the local chemical potential and $T({\bf
r},t)$ the local temperature.  The integrand in (8.4) can be rewritten as
${\partial s_{eq} \over \partial t} = - {\partial \over \partial {\bf r}}
\cdot {\bf j}_s ({\bf r},t) + \sigma({\bf r},t)$, where ${\bf j}_s$ is the
entropy flux and $\sigma({\bf r},t)$ is the local entropy production.  The
latter can be written in a form similar to (4.1) in terms of the full
pressure tensor ${\bf P}({\bf r},t)$, heat flux ${\bf J}({\bf r},t)$, etc.,
and is non-negative by the second law, see [1, l7, 34].  We thus find,
$$
\dot S_h(t) + \int_{\rm Surface} {\bf j}_s({\bf r},t)\cdot {\bf ds}  = \int_{\rm Volume} \sigma({\bf r},t)\, d{\bf r} \equiv
R_l(t)  \geq
0,\eqno(8.5)
$$

For an {\it isolated} macroscopic system the integral of ${\bf j}_s$ over
the surface vanishes and the rate of change of $S_h(t)$ is then given by
$R_l(t)$.  We can thus interpret \ $R_l$ as the rate at which $S_B$ is
produced inside the system when it is in LTE.  (LTE was implicitly assumed
in sections 3 and 4a.)  Furthermore, since the system is isolated we expect
it to approach, as $t \to \infty$, a uniform global equilibrium state, with
$S_h \to S_{\rm eq} = \left ({\rm Volume}\right )\cdot s_{\rm eq} (\bar n,
\bar e^\prime)$ and $R_l \to 0$.

Consider next the hydrodynamic description of an ``open'' macroscopic
system subjected to external forces and able to exchange heat through its
surface with $k$ heat baths maintained at temperatures $T_{\alpha}, \alpha
= 1,\ldots,k$.  (In the shear flow case we could have the walls at
different temperatures.)  We will now have, using (8.5) for the time
derivative of the system's hydrodynamic entropy, assuming that the part of
the system in contact with the $\alpha {\rm th}$ heat bath has a
temperature very close to $T_{\alpha}$, (we are considering for simplicity
the ``no slip'' case),
 
$$
\dot S_h(t) = 
R_l(t) - \sum J_\alpha (t) / T_\alpha.\eqno(8.6)
$$
Here $J_\alpha$ is the heat flux from the system to the $\alpha$th heat
bath, i.e.\ $J_\alpha/T_\alpha$ is the net hydrodynamic entropy flux
leaving the system through that part of the surface which is at temperature
$T_\alpha$.  Thus $J_\alpha / T_\alpha$ is analogous to the equilibrium
relation $dQ = dS/T$, extended to quasi-stationary processes; see in
particular chapter 14 in Balian's book [1].  Note that, in accordance with
(8.2), the external mechanical forces do not contribute directly to the
entropy change as they do not, by Liouville's theorem, change the phase
space volume $\Gamma (\cal M)$ available to the macro state $\cal M$.
 
When the differences in the $T^\prime _\alpha$s and the magnitude of the
external forces are not too large we expect this system to come to a
stationary LTE state in which $\dot S_h$ vanishes, with
$$
R_l \to \bar R_l = 
\sum \bar J_\alpha/T_\alpha;\eqno(8.7)
$$
 in accord with (4.2), for laminar shear flow.
 
Let us investigate now the statistical microscopic description of these
macroscopic situations.  Following the usual procedures of statistical
mechanics for macroscopic systems, we represent their macroscopic state,
${\cal M}$, at time $t$, by a suitable ensemble density $\mu(X,t)$.  This
$\mu$ is assumed to have the property that the hydrodynamic variables
${\cal M}_{\mu_t} = \{n, {\bf u}, e\}$ obtained as expectation values with
respect to $\mu(X,t)$ of the corresponding phase space functions are
sharply defined (very little dispersion with respect to $\mu$), vary slowly
in space and time on the microscopic scales and evolve, on the appropriate
macroscopic scales, according to the hydrodynamic equations. We will
associate to $\mu(X,t) \equiv \mu_t(X)$ the local equilibrium ensemble
density $\nu(X|{\cal M}_{\mu_t})$ and, with some abuse of notation, shall
set $\nu(X|{\cal M}_{\mu_t}) = \nu_t(X)$.  N.B.\ While we assume that our
system is in LTE, we are {\it not} assuming here that $\mu_t = \nu_t$.
 
\setcounter{equation}{7}
 
We now take the Gibbs entropy $S_G$ of the ensemble density $\mu_t$ defined
in (4.4), and split it into two parts,
\begin{eqnarray}
S_G(\mu_t) \equiv S_G(t) &\equiv& -\int \mu_t(X) \log \mu_t(X)\, dX \nonumber\\
       &=& -\int \mu_t(X) \log[\mu_t(X)/\nu_t(X)]\, dX 
        + \int \mu_t(X) \log \nu_t(X)\, dX
\end{eqnarray}
The first term on the right side of (8.8) is the relative entropy of
$\mu_t$ with respect to $\nu_t$, (which is negative, or zero, by convexity
of $x \log x$).  To evaluate the second term we use the fact that, by the
definition of $\nu$, as a locally grand-canonical ensemble density, $\log
\nu(X)$ is a function of $X$ whose average is expressible in terms of
${\cal M} = \{n,{\bf u}, e\}$.  Now, by the definition of $\nu_t$ these are
the actual hydrodynamical variables in our system so that the second term
is equal to $\int \nu_t(X) \log \nu_t(X) dX$.  Hence the second term in
(8.8) is, using (8.2), equal to $S_h(t)$, and we therefore have, that
$$
S_G(t) = S_G(\mu_t|\nu_t) + S_h(t)\eqno(8.9)
$$
Eq. (8.9), which is valid for all macroscopic systems in LTE, gives an
important connection between $S_G$ and $S_h$ for such systems.
 
For an isolated system with a given Hamiltonian $H$, we write $\dot X =
{\cal F}_H(X)$ for the flow in the phase space.  The ensemble density
$\mu(X,t)$ then evolves according to the Liouville equation,
$$
{\partial \mu(X,t) \over \partial t} = {\cal L}_H\, \mu = -{\rm div}(\mu
{\cal F}_H) = -\nabla \mu \cdot {\cal F}_H,\eqno(8.10)
$$
  subject to some
initial condition $\mu(x,0) = \mu_0(X)$.  (We think of $X, {\cal F},
\nabla$ as  $(2dN)$-dimensional vectors in the phase space and in the last
equality have used the fact that ${\rm div} {\cal F}_H = 0$.)  As was
already known to Gibbs, and is proven in almost every text book on
statistical mechanics, $S_G(t)$ is constant in time for an isolated system.
Hence, using (8.9)
$$
\dot S_G(t) = \dot S_G(\mu_t|\nu_t) + \dot S_h(t) =
0\eqno(8.11)
$$
or
$$
\dot S_G(\mu_t|\nu_t) = -\dot S_h(t) = -R_l(t)\eqno(8.12)
$$
Note that in going from (8.11) to (8.12) we are glossing over the
difference between microscopic and macroscopic time scales, see [11, 25,
34].  In the same spirit we will also have that $R_l(t) = \langle \tilde
R_l \rangle$ where $R_l$ is a suitable microscopic function and the average is
with respect to $\mu_t$, not $\nu_t$.
 
As $t \to \infty$, we expect that $\mu_t$ and $\nu_t$ will both
``approach'' the Gibbs distribution $\mu_{\rm eq}$ appropriate for a
macroscopic system in equilibrium with a uniform temperature and density.
Note however, that while $S_h(t)$ approaches $S_{\rm eq}$, the Gibbs
entropy, $S_G (\mu_t)$, being constant, clearly does not.  Hence if
$\mu_0(X)$ is not an equilibrium state, the limit, as $t \to \infty$, of
$S_G(\mu_t|\nu_t)$ will be $S_G(0) - S_{\rm eq}$, which is negative and can
be large, i.e.\ extensive, indicating the ``entanglement'' of $\mu_t(X)$
necessary to keep $S_G$ constant.
 
An explicit example where $S_G(\mu_t|\nu_t)$ can be shown to be extensive
for large $t$, occurs for non interacting point particles moving among a
periodic set of fixed scatterers, the periodic Lorentz gas or Sinai
billiard, which are started at $t=0$ in a product measure with the same
energy (speed) and a macroscopically {\it nonuniform} density.  The density
then evolves on the macroscopic scale according to the diffusion equation,
see [11, 25], and references there.  This leads to a uniformization of the
density and an increase in $S_h(t)$ while $S_G(t)$ remains constant.  The
entanglement of $\mu_t(X)$ corresponds here to the build up of correlations
between the positions and velocities of the particles in the system as it
evolves towards a spatially uniform state.  (The lack of interaction
between the particles makes this system a bit unphysical.)
 
We turn now to the statistical mechanics of open systems, e.g.\ a fluid
with shear flow, whose macroscopic behavior is described by the
compressible Navier-Stokes eqs.  Modeling such a system microscopically as
one driven by stochastic thermal reservoirs and some external forces,
e.g.\ rough walls with constant temperatures and velocities, as in section
4b, the time evolution of $\mu(X, t)$ will now be given by [4,5]
$$
{\partial \mu \over \partial t} = \left ({\cal L}_H + {\cal L}_{\rm ex} + \sum
{\cal K}_\alpha\right )\mu \eqno(8.13)
$$
Here ${\cal K}_\alpha$ represents the stochastic effect of the $\alpha$th
reservoir which tries to bring the system to equilibrium with a temperature
$T_\alpha$, and ${\cal L}_{\rm ex} \mu = -{\cal F}_{\rm ex} \nabla \mu$,
assuming that the external force ${\cal F}_{\rm ex}$ is (phase-space)
divergence free.  The Gibbs entropy $S_G(t)$ will no longer be constant in
time so (8.12) will no longer hold.  Assuming the system to be in LTE, we
shall have instead of (8.12) the behavior given in (8.6), with
$$
\dot S_h(t) = \langle  \tilde R_l \rangle - \sum \langle
\tilde J_\alpha \rangle / T_\alpha\eqno(8.14)
$$
where $\langle \cdot \rangle$ is the average with respect to $\mu(X,t)$ of
phase space functions $\tilde R_l$ and $\tilde J_\alpha$; $\tilde J_\alpha$
specifies the rate at which energy flows from the system to the $\alpha$th
reservoir.  As in (8.6) we are assuming here that in our system, the
spatial region in contact with the $\alpha$th reservoir is itself at a
temperature very close to $T_\alpha$.  (Compare (8.14) with (4.3) where
there is no assumption of LTE; ${\cal R}$ there corresponds to $\dot
S_G(\mu_t | \nu_t) + R_l$).
 
We may assume that our ensemble density $\mu(X,t)$, evolving according to
(8.13) will, in contrast to what happens in an isolated system, approach a
smooth stationary density $\bar \mu(X)$ while $\nu_t(X) \to \bar \nu(X) =
\nu(X|{\cal M}_{\bar \mu})$ [4,5,10]. 
Thus for $t \to \infty$, $S_h(t)$ will approach the hydrodynamic entropy
$\bar S_h$ in the MSNS, which according to (8.2) is given by
$$
\bar S_h = -\int \bar \nu(X) \log \bar \nu (X)\, dX.\eqno(8.15)
$$
Setting $\dot S_B(t) = 0$, in (8.14), and letting $\bar R_l$ be the
internal entropy production in the stationary state, we will have
$$
\bar R_l = 
\sum \langle \tilde J_\alpha \rangle_{\bar \mu} / T_\alpha.\eqno(8.16)
$$
Comparing (8.16) with (8.7) we can identify the average of $\tilde
J_\alpha$ with respect to $\bar \mu$, $\langle \tilde J_\alpha
\rangle_{\bar \mu}$, with the hydrodynamic steady state energy flux 
$\bar J_\alpha$, which, for a system in LTE, should be just $\langle \tilde
J_\alpha \rangle_{\bar \nu}$.  Since we further expect that the
stochasticity of the reservoirs will make the stationary $\bar \mu$ a
``smooth'' density, independent of the initial $\mu_0$, we should also have
here
$$
\dot S_G(t) \to 0, ~~~{\rm and}~~~ S_G(\mu_t |\nu_t) \to S_G(\bar \mu |
\bar \nu)~~~ {\rm for}~~~ t \to \infty \eqno(8.17)
$$
 
\setcounter{equation}{20}
 
Let us turn now to the microscopic modeling of the open macroscopic system
by ``thermostatted'' deterministic forces, ${\cal F}_{\rm ts}(X)$, which
conserve the energy but whose divergence does not vanish, e.g.\ the Maxwell
demon boundary drives discussed in this paper.  For the sake of simplicity
we shall treat ${\cal F}_{\rm ts}(X)$ as if it was smooth, but think of it
as acting only in the vicinity of the boundaries; the transition to a
boundary term should then be possible, see Appendix.  The time evolution of
a microstate $X$ will now be given by the deterministic equation
$$
\dot X = {\cal F}_H(X) + {\cal F}_{\rm ts}(X),\eqno(8.18)
$$
and that of $\mu(X,t)$ by
$$
{\partial \mu \over \partial t} = -{\rm div}[({\cal F}_H + {\cal
F}_{\rm ts})\mu].\eqno(8.19)
$$
  Taking the time derivative of $S_G (t)$, we get, as in (4.6), 
$$
\dot S_G(t) = \int \mu_t(X)\, {\rm div}\, {\cal F}_{\rm ts}(X)\, dX \equiv -M(t)\eqno(8.20)
$$
We also have, using (8.8) and (8.9), for systems in LTE, that 
\begin{eqnarray}
\dot S_h(t)  &=&  {d \over dt} \int \mu_t(X) \log \nu_t(X)\, dx
         = - \int {\cal F}_H(\nabla \cdot \mu_t) \log \nu_t\, dx\nonumber\\
    &-& \int {\rm div}({\cal F}_{\rm ts}) \mu_t \log \nu_t\, dx
   -\int \mu_t(X) {\partial \over \partial t} \log \nu_t(X)\, dX
\end{eqnarray}
 
In the last term on the rhs of (8.21) we can replace $\mu_t$ by $\nu_t$
since the time derivative of $\log \nu_t$ at a fixed $X$, involves only
phase space functions whose expectation corresponds to the hydrodynamic
variables.  After this replacement the last term becomes ${d \over dt}\int
\nu_t(X)
\, dX$,  which vanishes by normalization.  We are thus left with
\begin{eqnarray}
\dot S_h(t) = &-& \int ({\cal F}_H \cdot \nabla \mu_t) \log \nu_t\, dX\nonumber\\
&+& \int \mu_t {\cal F}_{\rm ts} \cdot \nabla \log \nu_t\, dX
\end{eqnarray}
 
We now want to argue that for ``smooth'' thermostatted forces we can again
replace $\mu_t$ by $\nu_t$ in the last term in (8.22), obtaining
\begin{eqnarray}
\int \nu_t {\cal F}_{\rm ts} \cdot \nabla \log \nu_t\, dX &=& -\int {\cal F}_{\rm ts}
\cdot \nabla \nu_t\, dX\nonumber\\
&=& \int \nu_t\, {\rm div}{\cal F}_{\rm ts}\, dX \equiv -M_l(t)
\end{eqnarray}
 
We also argue that the first term on the right side of (8.22) which is the
only term present in an isolated system, is just $R_l(t)$.  Accepting these
``arguments''  we finally get for systems with thermostatted forces in LTE, 
$$
\dot S_h(t) = R_l(t) - M_l(t),\eqno(8.24)
$$
 
We note that (8.24) gives a decomposition of the rate of change of $S_h$
into an internal part, $R_l$, coming from the dissipative fluxes inside the
system, and a thermostatted part, $-M_l$, coming from the thermostat forces
${\cal F}_{\rm ts}$.  Comparing this with (8.4) and remembering that ${\cal
F}_{\rm ts}$ does not change $n$ or $e$, $M_l$ will contribute only to the
last term there, $-\int {1 \over T}{\partial \over \partial t}({1
\over 2} nu^2)d{\bf r}$.  In fact, $M_l$ corresponds to the surface term in
(8.5), which for the models considered in this paper is equivalent to the
second term on the right side of (8.6): it represents the production (or
reduction) of $S_h$ due to the conversion of thermal into directed energy
by the Maxwell demons producing the ${\cal F}_{\rm ts}$ at the boundary.
 
Interestingly, the contribution of the ${\cal F}_{ts}$ to $\dot S_h$,
$-M_l(t)$, is just the rate of change of $S_G(t)$ {\it when} $\mu_t =
\nu_t$.  It is tempting to try and bypass the formal manipulations leading
to (8.24) and derive that relation more directly from the definition of
$S_B({\cal M})$ in (8.3) as $\log \Gamma({\cal M})$, but we have not
succeeded in doing this so far.

Waiting now for a time $\bar t$ which is long enough for the system to
become approximately stationary on the macroscopic level, we would have
$\dot S_h(t) \approx 0$, for $t \geq \bar t$, and thus
$$
\bar R_l = \bar M_l\eqno(8.25)
$$
Eq. (8.25) explains the equality between (5.12) and (5.15).  {\it If} it is
further true that ${\rm div}\dot{X}$ is sufficiently smooth for its
average to be well approximated by $M_l$, then we would have $\bar M =
\bar M_l = \bar R_l$.  This appears to be the case in many situations,
including the Maxwell demon boundaries considered in this paper, where
${\rm div} \dot X$ is an additive function of the coordinates and
velocities of each particle so that its average, $M$, depends only on the
one particle distribution function at the wall.  This leads to the
equality, $\bar M \approx \bar R_l$, which we observe in our simulations.
We actually expect the equality (8.25) to hold for general thermostatted
SNS {\it as long as} LTE {\it holds} in the SNS.  It should in particular
hold for macroscopic fluids, driven by rules (2.3) or (2.4) even when the
flow is no longer laminar.  We hope to test this via simulations on large
systems.

\medskip
 
\noindent {\bf Acknowledgments} 
During the course of this work we have benefited from discussions with many
colleagues.  In particular, we thank F. Alexander, H. van Beijeren,
F. Bonetto, E. Carlen, O. Costin, J. Erpenbeck, G. Eyink, G. Gallavotti,
S. Goldstein, W. Hoover, J. Koplik, Ya. Sinai, H. Spohn and H.T. Yau for
various helpful suggestions. We have also received a preprint by
Ch. Dellago and H.A Posch [37] who compute the Lyapunov exponents for our
$b$ and $c$ models.  Research was supported in part by AFOSR Grant
AF--92--J--0115, NSF Grants DMR 92-13424 and DMS-9401417, and the Faculty
Research Grant at the University of Alabama at Birmingham.
 
\renewcommand{\theequation}{\Alph{section}.\arabic{equation}}
 
\section*{Appendix. Special reflection rules}
\setcounter{section}{1}
\setcounter{equation}{0}
 
It seems reasonable to expect that time reversible reflection rules, like
rule b, can be obtained as limits of the usual smooth Gaussian
thermostatted dynamics [7, 8].  We describe here a very simple example of
such a limit which also provides an illustration of the relation between
$\bar M_l$ and $\bar R_l$, discussed in section 8.
 
Let a constant oblique force ${\bf F}=(E\cos\beta,-E\sin\beta )$ act in the
half-plane $y\geq L/2$, above the box, and a symmetric force ${\bf
F}=(-E\cos\beta,E\sin\beta)$ act below the box, in the area $y\leq
-L/2$. Here $0 < \beta < \pi/2$ is a fixed parameter, and $E>0$ is very
large, `almost infinite'. Such a force effectively replaces the walls.  The
force is accompanied by a Gaussian constraint which keeps the kinetic
energy of the particle fixed [7, 8].  The motion of a single particle in
the region $y\geq L/2$ is then governed by the equations,
$$
     \dot{x} = v_x, ~~~~ 
     \dot{y} = v_y
$$
and $\dot {\bf v} = {\cal F}_{ts}$, is given by 
 
\begin{eqnarray}
     \dot{v}_x &=& E\cos\beta-\alpha v_x\nonumber\\
     \dot{v}_y &=& -E\sin\beta-\alpha v_y,\nonumber
        \label{4}
\end{eqnarray}
where
\be
       \alpha = \frac{Ev_x\cos\beta-Ev_y\sin\beta}{v_x^2+v_y^2}
          \label{alpha}
\ee
so that $v^2$ is constant.   Symmetric equations hold 
for $y \leq -L/2$.
 
Taking now the limit $E \to \infty$ this model reduces to our Maxwell demon
 model (2.1) with a specific function $f=f_\beta$.  To obtain that, we take
 advantage of the conservation of kinetic energy and rewrite the system
 (A.1) as
\begin{eqnarray}
     \dot{x} &=& v\cos\theta\nonumber\\
     \dot{y} &=& v\sin\theta
        \label{3}\\
     \dot{\theta} &=& -\frac{E}{v}\sin(\theta+\beta)\nonumber
\end{eqnarray}
where $\theta=\tan^{-1}(v_y/v_x)$ is the angle between the velocity vector
$(v_x,v_y)$ and the $x$ axis. The last equation in (\ref{3}) is independent
of the first two, and has an implicit solution given by
\be
   \ln\left [ \frac{1+\cos(\theta+\beta)}{1-\cos(\theta+\beta)}\right ]
       =2\,\frac {Et}{v}
       \label{thetat}
\ee
It is, however, more convenient to differentiate $\theta$ with respect 
to the height, $h=y-L/2$, yielding 
\be
   \frac{d\theta}{dh}=-\frac{E}{v^2}\cdot \frac{\sin(\theta+\beta)}{\sin\theta}
       \label{thetadh}
\ee
An implicit solution of this equation is
\be
   (\theta+\beta)\cdot\cos\beta-\sin\beta\cdot\ln|\sin(\theta+\beta)|
          =-\frac{Eh}{v^2}
       \label{thetah}
\ee
Since $h=0$ both as the particle enters the force zone and as it leaves it,
the relation between the incoming angle, $\varphi=\theta_{\rm in}$ and the
outgoing angle, $\psi=-\theta_{\rm out}$, is determined by the equation
\be
     F_\beta(-\psi) = F_\beta(\varphi)
       \label{FF}
\ee
where
\be
   F_{\beta}(\theta)
     :=(\theta+\beta)\cdot\cos\beta-\sin\beta\cdot\ln|\sin(\theta+\beta)|
       \label{Fbeta}
\ee
 
These equations do not contain the force strength, $E$. By taking the limit
$E\to\infty$ we simply assure that the particle leaves the force zone
instantly, the moment it enters it. Therefore, the infinite force acts like
a wall at which the particle gets reflected, and the outgoing angle $\psi$
is related to the incoming angle $\varphi$ by equation (2.1) with
$f_{\beta}(\varphi)=-F_{\beta}^{-1} F_{\beta}(\varphi)$.

For $E$ very large the time any particle spends in the force zone is
extremely short, so we can neglect possible collisions between particles
while one of them is in that zone.  In fact if a particle enters the force
zone at time $s$ and leaves it at time $s+\tau$ we can assume that there is
no other particle in the force zone between $s$ and $s+\tau$.  Under these
conditions the analysis of section 8 takes on a particularly simple form.
 
Assuming that the stationary state is one of LTE, the one particle
distribution {\it in the region of the field} is a local Maxwellian where
$\Psi({\bf r}, {\bf v})$
\be
\Psi({\bf r}, {\bf v}) = n({\bf r}) (2\pi T_w)^{-1} 
   \exp\{-({\bf v} - {\bf v}_w)^2/2T_w\}
\ee
Using the definition of ${\bar M}_l$ in (8.23) with ${\cal F}_{\rm ts}$,
given in (A.1), we obtain
\begin{eqnarray}
\bar M_l &=& - \int d{\bf r} \int ({\partial \over \partial {\bf v}} 
    {\cal F}_{\rm ts}) \Psi({\bf r}, {\bf v})\, d{\bf v}\nonumber\\
    &=& -\int d{\bf r} \int {({\bf v} -{\bf v}_w) \over T_w} 
    \cdot {\cal F}_{\rm ts} \Psi({\bf r}, {\bf v})\, d{\bf v}\nonumber\\
    &=& {{\bf v}_w \over T_w} \int d{\bf r} \int {\dot {\bf v}} 
    \Psi({\bf r}, {\bf v})\, d{\bf v}
         \label{threeeq}
\end{eqnarray}
 
In going from the second to the third equality on the right side of
(\ref{threeeq}) we have used the conservation of energy by ${\cal F}_{\rm
ts}$, i.e.  $\dot{\bf v}\cdot {\cal F}_{\rm ts}$ = 0.  The final term is
easily recognized as corresponding to the last term in (8.4), giving the
entropy production due to ${\cal F}_{\rm ts}$.  It becomes equal to $\bar
R_l$ in (4.2) in the limit $E \to \infty$, giving $\bar M_l = \bar R_l$,
for such systems in LTE.

\end{document}